\documentclass[a4paper,11pt]{article}
\pdfoutput=1
\usepackage{amsmath}
\usepackage{jheppub}
\usepackage{amssymb}
\usepackage{graphicx}
\usepackage[T1]{fontenc}
\usepackage[utf8x]{inputenc}
\usepackage{color}

    \title{On the coverage of neutralino dark matter in coannihilations at the upgraded LHC}

    \author[a,b,c]{Murat Abdughani,}
    \author[a]{and Lei Wu}

    \affiliation[a]{Department of Physics and Institute of Theoretical Physics, Nanjing Normal University, Nanjing, 210023, China}
    \affiliation[b]{CAS Key Laboratory of Theoretical Physics, Institute of Theoretical Physics, Chinese Academy of Sciences, Beijing 100190, China}
    \affiliation[c]{School of Physics, University of Chinese Academy of Sciences, Beijing 100049, China}

    \emailAdd{mulati@mail.itp.ac.cn}
    \emailAdd{leiwu@njnu.edu.cn}

    \abstract{In the supersymmetric models, the coannihilation of the neutralino DM with a lighter supersymmetric particle provides a feasible way to accommodate the observed cosmological DM relic density. Such a mechanism predicts a compressed spectrum of the neutralino DM and its coannihilating partner, which results in the soft final states and makes the searches for sparticles challenging at colliders. On the other hand, the abundance of the freeze-out neutralino DM usually increases as the DM mass becomes heavier. This implies an upper bound on the mass of the neutralino DM. Given these observations, we explore the HE-LHC coverage of the neutralino DM for the coannihilations. By analyzing the events of the multijet with the missing transverse energy ($E^{miss}_T$), the monojet, the soft lepton pair plus $E^{miss}_T$, and the monojet plus a hadronic tau, we find that the neutralino DM mass can be excluded up to 2.6, 1.7 and 0.8 TeV in the gluino, stop and wino coannihilations at the $2\sigma$ level, respectively. However, there is still no sensitivity of the neutralino DM in stau coannihilation at the HE-LHC, due to the small cross section of the direct stau pair production and the low tagging efficiency of soft tau from the stau decay.
    }

\begin{document}

    \maketitle

\section{Introduction}
The presence of dark matter (DM) in the Universe has been established by versatile astrophysical and cosmological observations. However, its nature still remains a mystery. Weakly Interacting Massive Particles (WIMPs) are among the compelling dark matter candidates, whose mass is basically in the range from about 2 GeV~\cite{Lee:1977ua} up to several 100 TeV~\cite{Griest:1989wd} and interaction strengthes are of the order of weak coupling of the Standard Model. As the WIMP DM can naturally produce the measured thermal relic density, there have been many experiments devoting into the searches of WIMP DMs~\cite{Arcadi:2017kky}.

In the minimal supersymmetric standard model (MSSM), the lightest neutralino $\tilde{\chi}^0_1$ can serve as the WIMP DM candidate when the $R$-parity is conserved. According to the composition of $\tilde{\chi}^0_1$, it can be bino-like, wino-like, higgsino-like or the mixed state. If $\tilde{\chi}^0_1$ is wino-like or higgsino-like state, their annihilation rates are large so that their masses have to be in TeV region to saturate the observed DM relic density~\cite{ArkaniHamed:2006mb}. On the other hand, if $\tilde{\chi}^0_1$ is bino-like state, its interaction with the SM particles are weak, which usually leads to an overabundance of DM. Among ways to solve this problem, coannihilation of $\tilde{\chi}^0_1$ with a light sparticle is an interesting mechanism~\cite{Griest:1990kh}. Several coannihilation scenarios have been studied in supersymmetric models, such as the constrained MSSM (CMSSM) or mSUGRA, where the large negative value of $A_0$ drives the lighter stop or stau to be degenerate with the neutralino LSP, and the relic density of $\tilde{\chi}^0_1$ is brought into the range allowed by coannihilation with the stop or stau~\cite{Ellis:1998kh,Ellis:1999mm,Ellis:2001nx,Ellis:2014ipa,Desai:2014uha,Citron:2012fg,Han:2016gvr}. Besides, in the MSSM with non-universal gaugino masses or the vector-like extension of the MSSM, the gauginos can be the NLSP with a mass sufficiently close to that of the LSP so that gaugino coannihilation becomes important~\cite{BirkedalHansen:2002sx,Profumo:2004wk,Baer:2005jq,Feldman:2009zc,Ibe:2013pua,Harigaya:2014dwa,Ellis:2015vna,Ellis:2015vaa,Yanagida:2019evh}.

Coannihilating DM usually has some distinctive phenomenologies in DM experiments. For example, with the development of low noise technique and the increasing volume of detector, the sensitivity of DM direct detection experiment has been greatly improved. The null results produce strong constraints on various WIMP DM models. In coannihilation, due to the nature of bino, the neutralino DM weakly interacts with the quarks so that it can escape the bounds from the direct detection experiments. 

Besides, there have been a great efforts devoted to searching for sparticles at the LHC. The colored sparticles have been excluded up to TeV in simplified models. However, these exclusion limits in the compressed regions will become weak, and even vanish. It should be mentioned that various compressed SUSY mass spectra have been also studied in LHC experiments. For example, the mass difference of compressed electroweakinos (such as bino-wino coannihilation scenario) can be probed down to about 2 GeV by the current ATLAS search of two soft leptons plus missing transverse energy events~\cite{Aaboud:2017leg}. However, when the mass difference between the LSP and NLSP is of the order of the rho meson or below (but is still larger than about 400 MeV)~\footnote{If the mass splitting is less than about 400 MeV, the disappearing track search can be used to investigate are such almost degenerate electroweakinos~\cite{ATL-PHYS-PUB-2017-019}. }, there is still no limit at the LHC. It should be mentioned that the future hadron colliders are hardly to improved the sensitivity in these compressed regions because of the low reconstruction efficiency of soft leptons and contamination of high pile-up~\cite{ATL-PHYS-PUB-2018-031}.

In coannihilation, the mass difference between the neutralino DM and its coannihilating partner is usually small. Consequently, the light sparticles in coannihilation can be still consistent with the LHC data. Another interesting fact is that the abundance of neutralino DM in the freeze-out mechanism usually increases as the mass of the neutralino DM becomes heavier. This leads to the upper limits of the neutralino DM and its coannilating partner masses, which provides a guideline of searching for supersymmetry at colliders. Therefore, it is interesting to explore various coannihilation scenarios at the LHC and future colliders~\cite{Arnowitt:2006jq,Khotilovich:2005gb,Berggren:2015qua,Florez:2016lwi,Aboubrahim:2017aen,Drees:2012dd,Yu:2012kj,Kobakhidze:2015scd,Nath:2016kfp,Nagata:2015hha,Duan:2018rls,Nagata:2015pra,Nagata:2016hps}.

Beyond the LHC, the high-energy LHC (HE-LHC) is proposed to built on current LHC tunnel by upgrading to 16 T superconducting magnet~\cite{Abada:2019ono}. The HE-LHC is designed to operate at a center of mass energy $\sqrt{s}=27$ TeV, and to collect of the order of 15 ab$^{-1}$ of data during 20 years of operation, which can greatly extend the HL-LHC potential of accessing the mass ranges of new particles~\cite{CidVidal:2018eel,Atlas:2019qfx}. In Ref.~\cite{Han:2018wus}, the authors investigated for observability of TeV higgsino and wino-like neutralino DMs at the HE-LHC. 
In Ref.~\cite{Baer:2018hpb}, the authors studied the phenomenology of stop, gluino and higgsinos in natural SUSY at the LHC. Other studies of the discovery reach of supersymmetric particles at the HE-LHC can be found in~\cite{Aboubrahim:2018tpf,DiLuzio:2018jwd,Aboubrahim:2018bil,Han:2019grb,Aboubrahim:2019qpc,Aboubrahim:2019vjl}. In this paper, we will explore prospects for the coverage of neutralino dark matter coannihilation at the HE-LHC. We begin to identify the parameter space of stop, gluino, wino and stau coannihilation under the LHC and DM constraints in Section~\ref{section2}, and then perform the detailed Monte Carlo (MC) simulation for each coannihilation scenario at the HE-LHC in Section~\ref{section3}. Finally, we draw conclusions in Section~\ref{section4}.




	
\section{Coannihilating neutralino DM}\label{section2}

After the electroweak symmetry is broken in the MSSM, the mass matrix for neutralinos in bino($\tilde{B}$)-wino($\tilde{W}$)-higgsinos($\tilde{H}_{u,d}^0$) basis can be written as
\begin{eqnarray}
  \label{nmass}
  M_{\tilde{\chi}^0}=
\left(
   \begin{array}{cccc}
     M_1 &0&-\cos\beta \sin\theta_w m_Z&\sin\beta \sin\theta_w m_Z\\
     0& M_2&\cos\beta \cos\theta_w m_Z& \sin\beta \cos\theta_w m_Z\\
     -\cos\beta \sin\theta_w m_Z& \cos\beta \cos\theta_w m_Z&0&-\mu\\
     \sin\beta \sin\theta_w m_Z&-\sin\beta \cos\theta_w m_Z&-\mu&0\\
     \end{array}
 \right)
\end{eqnarray}
where $M_{1}$ and $M_{2}$ are $U(1)_Y$ and $SU(2)_L$ soft supersymmetry breaking mass parameters, respectively. $\mu$ is the higgsino mass parameter and $\theta_w$ is the weak mixing angle. We can diagonalize the Eq.~(\ref{nmass}) by a unitary $4\times 4$ matrices $N_{ij}$~\cite{Gunion:1984yn}, and then have the mass eigenstates $\tilde{\chi}^0_{1,2,3,4}$. When the $R$-parity is conserving, the lightest neutralino $\tilde{\chi}^0_1$ can play the role of the DM and provide the correct relic density by itself. However, if there exists other sparticles whose masses are nearly degenerate with $\tilde{\chi}^0_1$, the relic abundance of the neutralino DM is determined not only by its annihilation cross section, but also by the annihilation of these heavier sparticles. This case is, namely, coannihilation.
The effective coannihilation cross section can be written as~\cite{Baker:2015qna},
\begin{align} \label{eq:sigmaeff}
  \sigma_{\text{eff}} = \frac{g_{\tilde{\chi}^0_1}^2}{g_{\text{eff}}^2} \bigg\{ & \sigma_{\tilde{\chi}^0_1 \tilde{\chi}^0_1} + 2 \sigma_{\tilde{\chi}^0_1 \, \mathcal{P}} \frac{g_{\mathcal{P}}}{g_{\tilde{\chi}^0_1}} (1 + \Delta)^{3/2} \exp(-x \Delta) + \sigma_{\mathcal{P} \mathcal{P}} \frac{g_{\mathcal{P}}^2}{g_{\tilde{\chi}^0_1}^2} (1 + \Delta)^3 \exp(-2 x \Delta) \bigg\} \, .
\end{align}
Here $\Delta = (m_{\mathcal{P}} - m_{\tilde{\chi}^0_1})/m_{\tilde{\chi}^0_1}$ and $x = m_{\tilde{\chi}^0_1}/T$. The parameters $g_{\tilde{\chi}^0_1}$ and $g_{\mathcal{P}}$ are the numbers of degrees of freedom of DM and coannilating partner $\mathcal{P}$, respectively. The effective coupling $g_{\text{eff}}$ is given by,
\begin{equation}
g_{\text{eff}} = g_{\tilde{\chi}^0_1} + g_{\mathcal{P}} (1+ \Delta)^{3/2} \exp ( -x \Delta) \, .
\end{equation}
From Eq.~\ref{eq:sigmaeff}, it can be seen that the contributions of the terms including $\sigma_{\tilde{\chi}^0_1 \, \mathcal{P}}$ and $\sigma_{\mathcal{P} \mathcal{P}}$ can become important, even dominant over $\sigma_{\tilde{\chi}^0_1 \tilde{\chi}^0_1}$, when $\Delta$ is vanishing.

We will carry out our study of coannihilations in the simplified MSSM, where only relevant sparticles in each scenario are involved. Such a framework allows us to remain agnostic of the detailed UV-physics, yet still capture the feature of coannihilation. We scan the ranges of SUSY mass parameters  in gluino, stop, wino and stau coannihilations as following:
\begin{eqnarray}
  &&{\rm \textbf{Bino-Gluino:}} \quad 100~{\rm GeV} < M_{1,~3} < 3~{\rm TeV}, 1 < \tan\beta < 60,\nonumber\\
  &&{\rm \textbf{Bino-Stop:}} \quad 100~{\rm GeV} < M_{1,~QL3,~U3R} < 2~{\rm TeV}, |A_t| < 3~{\rm TeV}, 1 < \tan\beta < 60,\nonumber \\
  &&{\rm \textbf{Bino-Wino:}} \quad 100~{\rm GeV} < M_{1,~2} < 1~{\rm TeV}, 1 < \tan \beta < 60, \nonumber\\
  &&{\rm \textbf{Bino-Stau:}} \quad 100~{\rm GeV} < M_{1,~L3L,~E3R} < 3~{\rm TeV}, |A_\tau| < 3~{\rm TeV}, 1 < \tan \beta < 60.\nonumber
\end{eqnarray}
In above each scenario, we assume the CP odd Higgs mass $m_A$ and other soft SUSY breaking mass parameters as a common value $M_{SUSY}=5$ TeV, and take other irrelevant trilinear soft SUSY breaking parameter $A=0$. We calculate the DM relic density $\Omega_{\tilde{\chi}} h^2$ with \textsf{MicrOMEGAs}~\cite{Belanger:2013oya} and the Higgs mass with~\textsf{SUSY-HIT}~\cite{Djouadi:2006bz}. We require our samples to satisfy the $2\sigma$ bounds of the Planck value of DM relic density~\cite{Ade:2013zuv} and the measured Higgs mass within the range of $125 \pm 3$ GeV~\cite{Aad:2015zhl}. In addition, we impose the vacuum stability constraints in stop and stau coannihilations, since the large mixing in stop and stau sector may lead to the charge or color breaking~\cite{Chowdhury:2013dka,Kitahara:2013lfa,Duan:2018cgb}. Besides, we also consider the experimental 95\% C.L. exclusion limits from the null results of the LHC searches for gluino~\cite{Aaboud:2017vwy}, stop~\cite{Aaboud:2017phn} and wino~\cite{Aaboud:2017leg}, which are calculated by using $CL_s$ prescription with the expected and observed number of events at the experiment.

\begin{figure}[ht]
		\includegraphics[width=8cm,height=8cm]{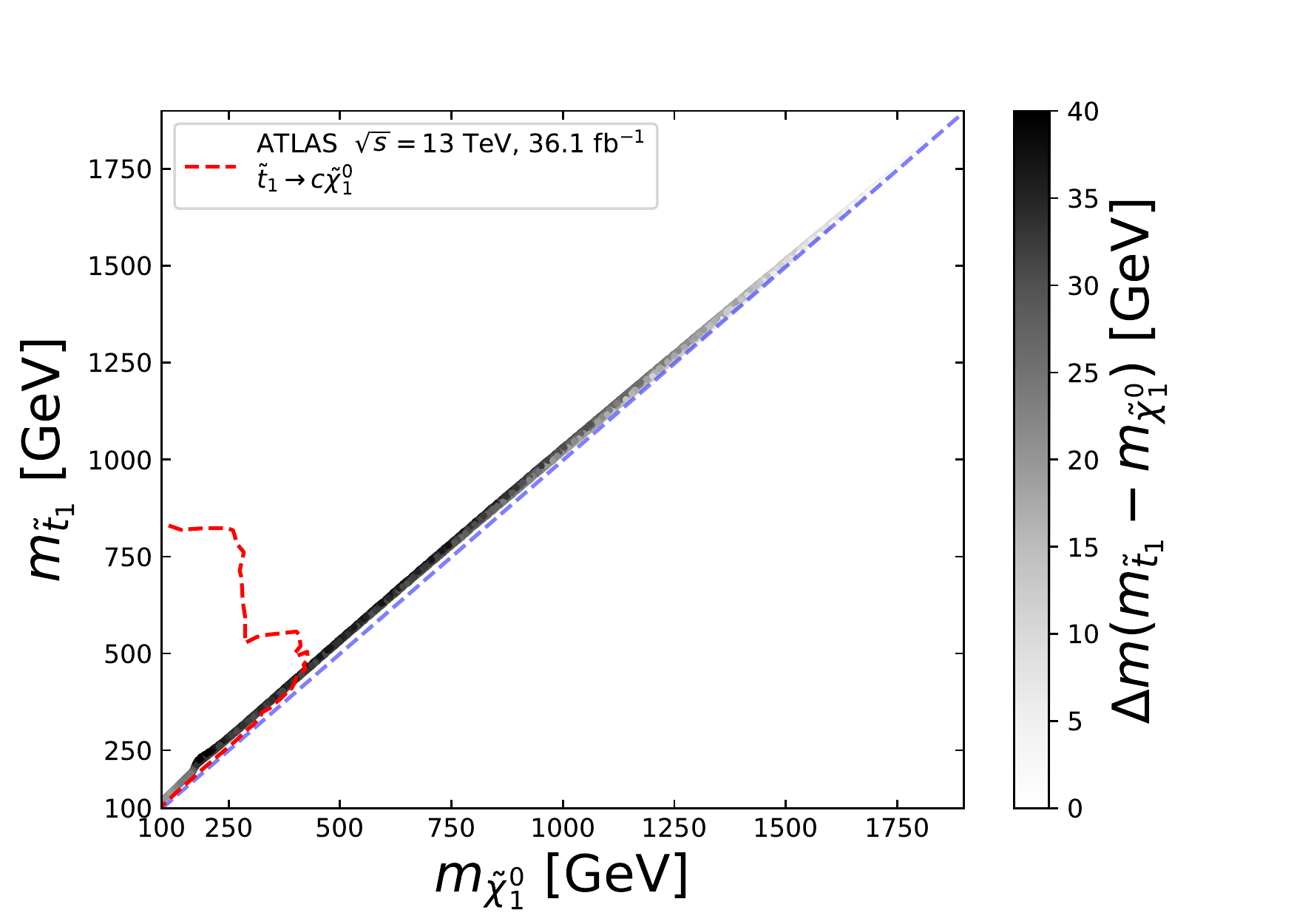}
        \includegraphics[width=8cm,height=8cm]{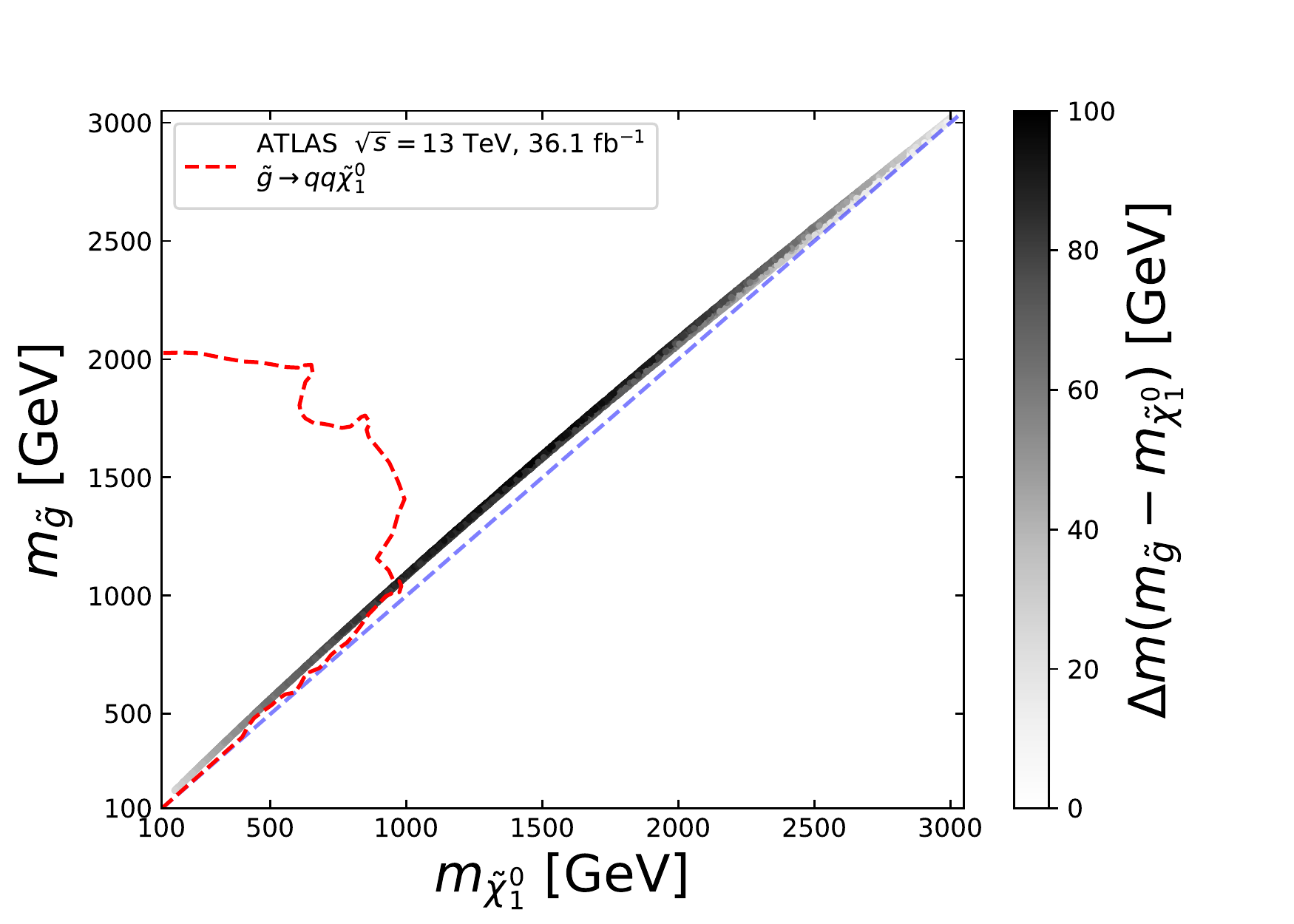}
        \includegraphics[width=8cm,height=8cm]{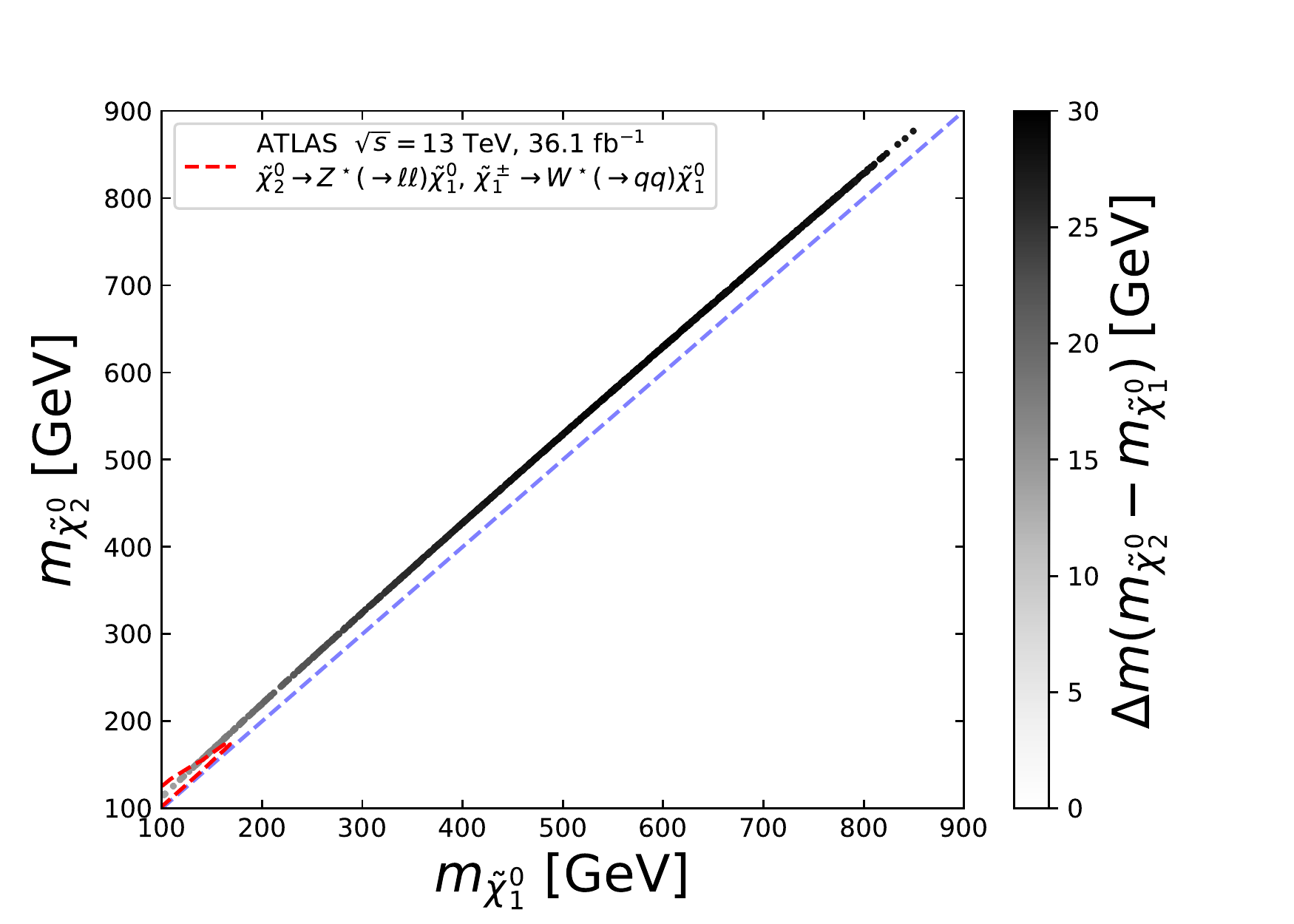}
        \includegraphics[width=8cm,height=8cm]{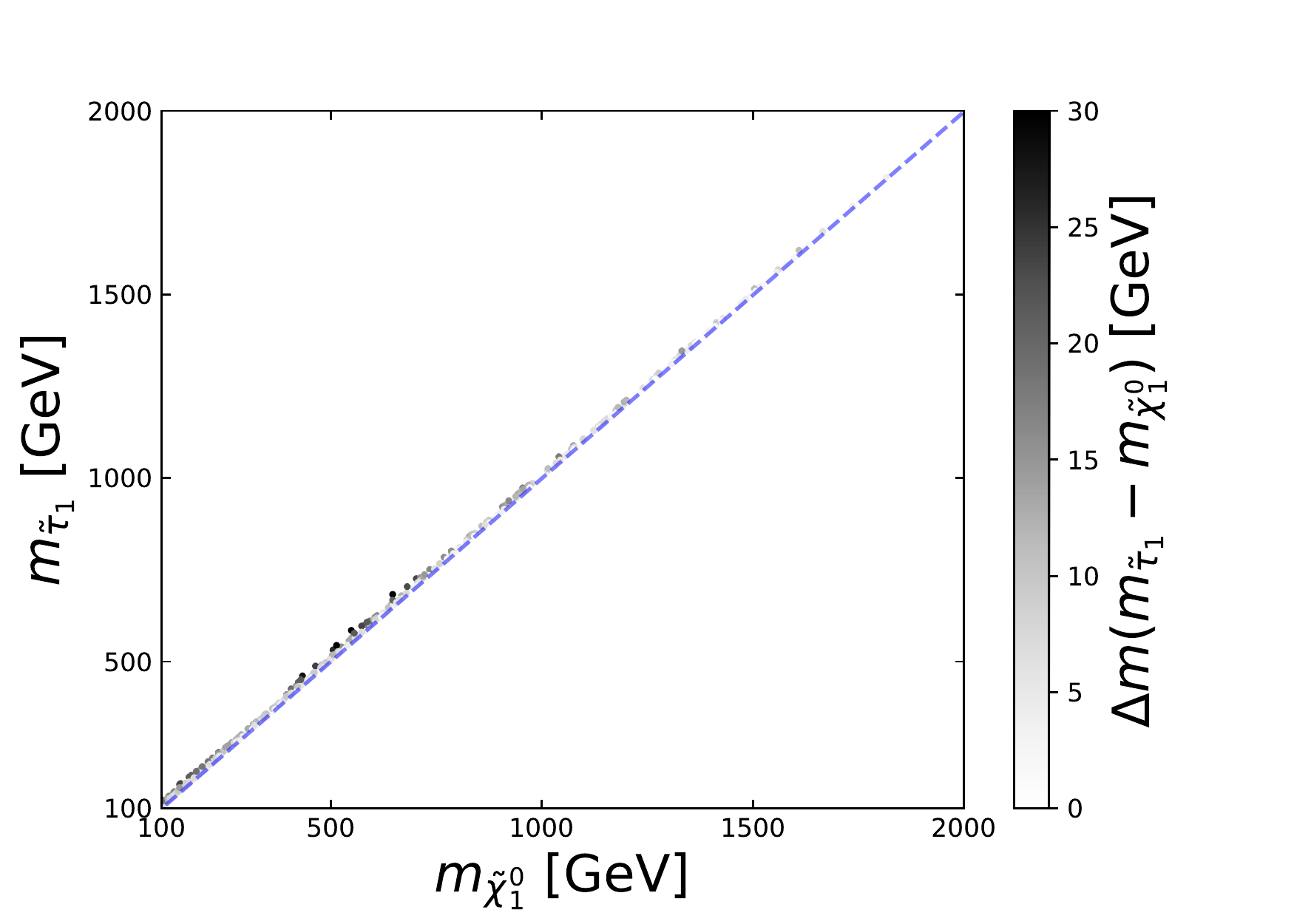}
		\caption{The samples that satisfy the observed DM relic density, the Higgs mass and vacuum stability conditions are projected on the plane of DM mass and its coannihilating partner mass for bino-stop (top-left panel), bino-gluino (top-right panel), bino-wino (lower-left panel) and bino-stau (lower-right panel) coannihilations. The colormap represents the mass difference of DM and coannihilating partner in each scenario. The red dashed lines denotes the 95\% C.L. exclusion limits from the null results of searching for gluino~\cite{Aaboud:2017vwy}, stop~\cite{Aaboud:2017phn} and wino~\cite{Aaboud:2017leg} at the LHC-13 TeV with the luminosity ${\cal L}=36.1$ fb$^{-1}$.}
		\label{fig:mass-range}
\end{figure}
In Fig.~\ref{fig:mass-range}, we show the mass ranges of the neutralino DM and its coannihilating partners for the surviving samples allowed by the DM relic density, the Higgs mass and the vacuum stability conditions for each coannihilation. As known, when the neutralino DM becomes heavy, the abundance of the neutralino DM will overclose the Universe. This leads to an upper limit of the neutralino DM mass. Among these scenarios, the gluino and stop interact strongly and extend the allowed mass range of the LSP for accommodating the correct relic density, while the wino and stau interact weakly and have to be lighter. From the Fig.~\ref{fig:mass-range}, we can see that the neutralino DM and stop masses in the stop coannihilation should be lighter than 1.9~TeV. The mass difference between the neutralino DM and the stop varies from 2~GeV to 40~GeV. Most of samples are right-handed stop, which will decay through the loop process $\tilde t_1 \to \tilde \chi_1^0 + c$ and four-body process $\tilde t_1 \to \tilde \chi_1^0 + b f \bar f'$. In the gluino coannihilation, the upper limit of the neutralino DM mass is about 3 TeV, which is much greater than that in other three scenarios, because the gluino-gluino annihilation has a larger cross section. For the same reason, the maximal mass difference between the neutralino DM and the gluino can reach about 100 GeV. Such a gluino mainly decays through the process $\tilde g \to \tilde \chi_1^0 q \bar q$. While for the wino coannihilation, the neutralino DM and the next-to-lightest neutralino $\tilde{\chi}^0_2$ masses are less than 900~GeV. Due to their small mass splittings, the coannihilating partners will dominantly decay to the SM particles via $\tilde\chi_2^0 \to Z^* \to \tilde{\chi}_1^0 f \bar f$ and $\tilde\chi_1^\pm \to W^{\pm*} \to \tilde \chi_1^0 f \bar f^\prime$. As for stau coannihilation, the main contribution to the relic density comes from stau-stau annihilation into $\tau^+\tau^-$ for light stau, and into $hh$, $ZZ$ and $W^\pm W^\mp$ for heavy stau. Besides, in Fig.~\ref{fig:mass-range}, we can also see that the null results of the LHC searches for sparticles have produced the lower bounds on gluino, stop and wino masses in coannihilation regions, that is, $m_{\tilde{t}_1}>400$ GeV, $m_{\tilde{g}}>1000$ GeV and $m_{\tilde{\chi}^0_2}>150$ GeV. While there is still no stronger limit on the stau mass from the LHC data than that from the LEP experiment.

Due to the Sommerfeld enhancement effects at low velocities, the coannihilation rates can be increased so that the upper bounds of the neutralino DM masses will be altered~\cite{Hisano:2004ds,Hisano:2006nn,Hryczuk:2010zi,Hryczuk:2011tq,Beneke:2012tg,Hellmann:2013jxa,Beneke:2014gja,Beneke:2014hja,Beneke:2016ync}. For example, the bino-like neutralino DM mass in stop coannihilation consistent with the observed DM abundance turns out to be several TeV~\cite{Ellis:2014ipa,Ellis:2018jyl}. Besides the Sommerfeld enhancement, the bound-state effects~\cite{Shepherd:2009sa,vonHarling:2014kha,Liew:2016hqo,Binder:2018znk,Fukuda:2018ufg} and the higher order perturbative corrections~\cite{Moroi:2006fp,Herrmann:2007ku,Hryczuk:2011vi,Bringmann:2015cpa,Klasen:2016qyz} can further increase the neutralino DM coannihilation rates and thus extend the neutralino DM mass that can give the observed DM relic density. Given the designed colliding energy of the HE-LHC, we do not include those two effects in our calculations because the following results of their masses reach will not be changed. Besides, it should be noted that the decay widths of the coannihilating partners will become small when their masses are very close the neutralino DM. In this case those sparticles will have a long life in the detector. We leave the detailed analysis of searching for the long lived sparticles in our future work.

\section{Prospects for coannihilations at the HE-LHC}\label{section3}

Next, we study the prospects of searching for these sparticles in coannihilations at the HE-LHC. The cross sections of our signal and background processes are calculated at LO. We simulated the signal and background events by the package \textsf{MG5\_aMC@NLO v2.6.1}~\cite{Alwall:2014hca} with the NN23LO1 PDF (Parton Distribution Function) set. 
Then the parton shower and hadronization are performed by the package \textsf{Pythia8.230}~\cite{Sjostrand:2014zea}. The jets are clustered by using the anti-$k_t$ algorithm~\cite{Cacciari:2008gp} with the distance parameter $R = 0.4$. We implement the fast detector simulation by the package \textsf{Delphes3.4.1}~\cite{deFavereau:2013fsa}. It should be noted that the parton-level events of $Z/W + jets$ are generated up to two jets that are matched to the parton shower using the MLM-scheme with merging scale $Q=60$ GeV. The event selections are carried out in the framework of \textsf{CheckMATE-2.0.26}~\cite{Drees:2013wra}.
We evaluate the statistical significance with the formula $Z=S/\sqrt{B}$, where $S$ and $B$ denote the signal and background yields respectively.

\subsection{Gluino coannihilation}

\begin{figure}[ht]
  \centering
  \includegraphics[width=12cm,height=8cm]{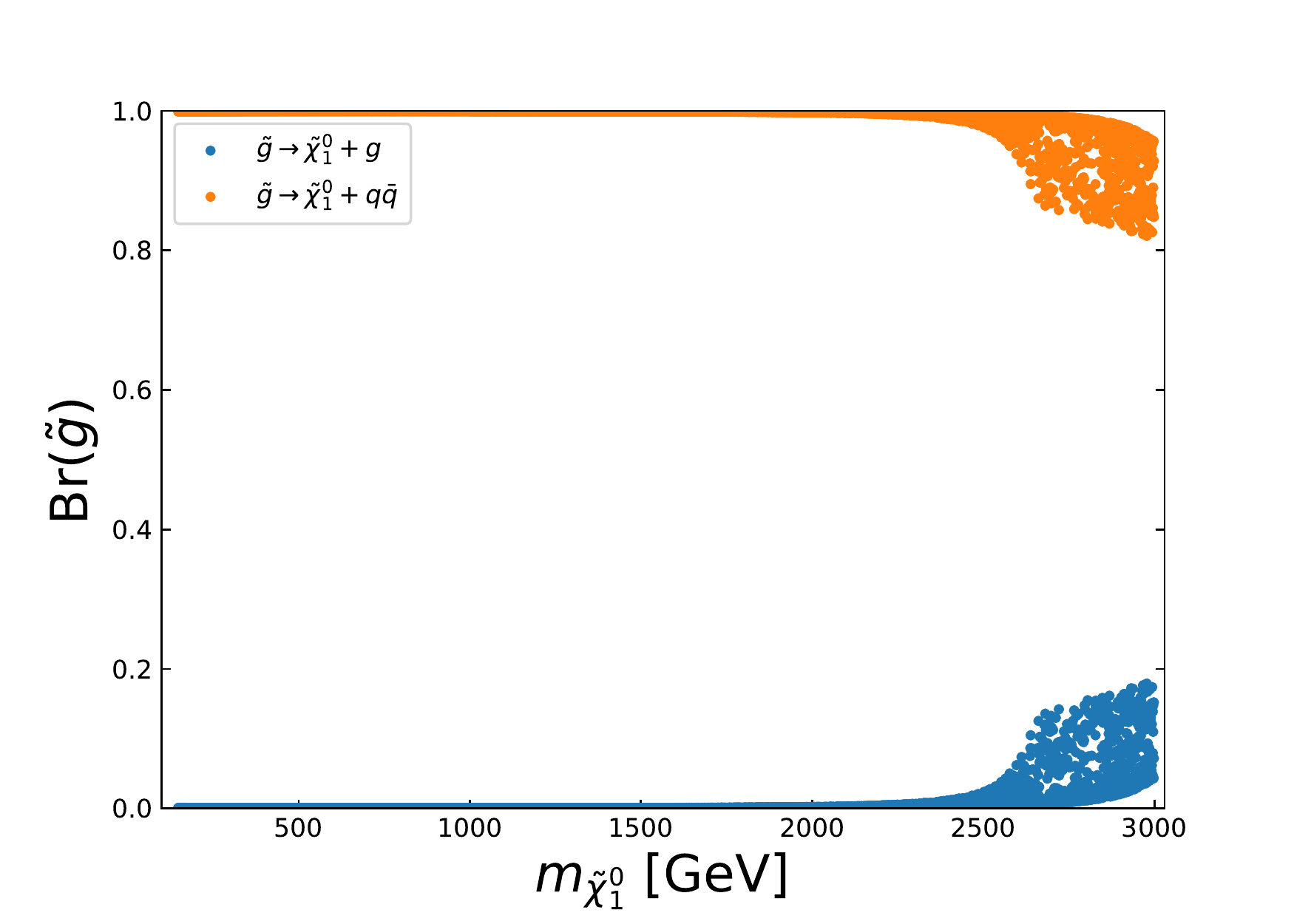}
  \caption{The branching ratio of $\tilde{g}$ in the gluino coannihilation.}
  \label{fig:br_gluino}
\end{figure}

\begin{figure}[ht]
  \centering
  \includegraphics[width=6cm,height=4cm]{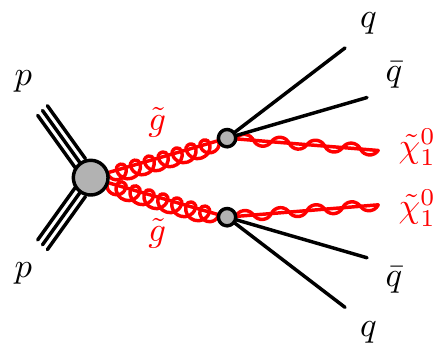}
  \caption{The schematic diagram of the production process $pp \to \tilde{g}\tilde{g}$ with the sequent decay $\tilde{g} \to q\bar{q}\tilde{\chi}^0_1$ at the HE-LHC.}
  \label{fig:feyn_gluino}
\end{figure}

The gluinos are mainly produced through two processes $gg \to \tilde{g}\tilde{g}$ and $q\bar{q} \to \tilde{g}\tilde{g}$ at the hadron colliders. In the gluino-neutralino coannihilation, the decay of the gluino is dominated by the three-body decay process $\tilde{g} \to q\bar{q}\tilde{\chi}^0_1$, and the subleading decay is the loop-induced process $\tilde{g} \to g \tilde{\chi}^0_1$ that is typically a few percent, which can be seen from Fig.~\ref{fig:br_gluino}. Therefore, we use multi-jet plus missing transverse energy events from the processs $pp \to \tilde{g}\tilde{g}$ with the sequent decay $\tilde{g} \to q\bar{q}\tilde{\chi}^0_1$ to probe the gluino coannihilation at the HE-LHC, as shown in Fig.~\ref{fig:feyn_gluino}. The main SM backgrounds include: $W(\rightarrow \ell v)$ + jets, $Z/\gamma^\star(\rightarrow \ell \bar \ell)$ + jets, $\gamma$ + jets, $t \bar t$, single top, and dibosons ($WW,\ WZ,\ ZZ$).

\begin{figure}[ht]
  \centering
  \includegraphics[width=7.5cm,height=9cm]{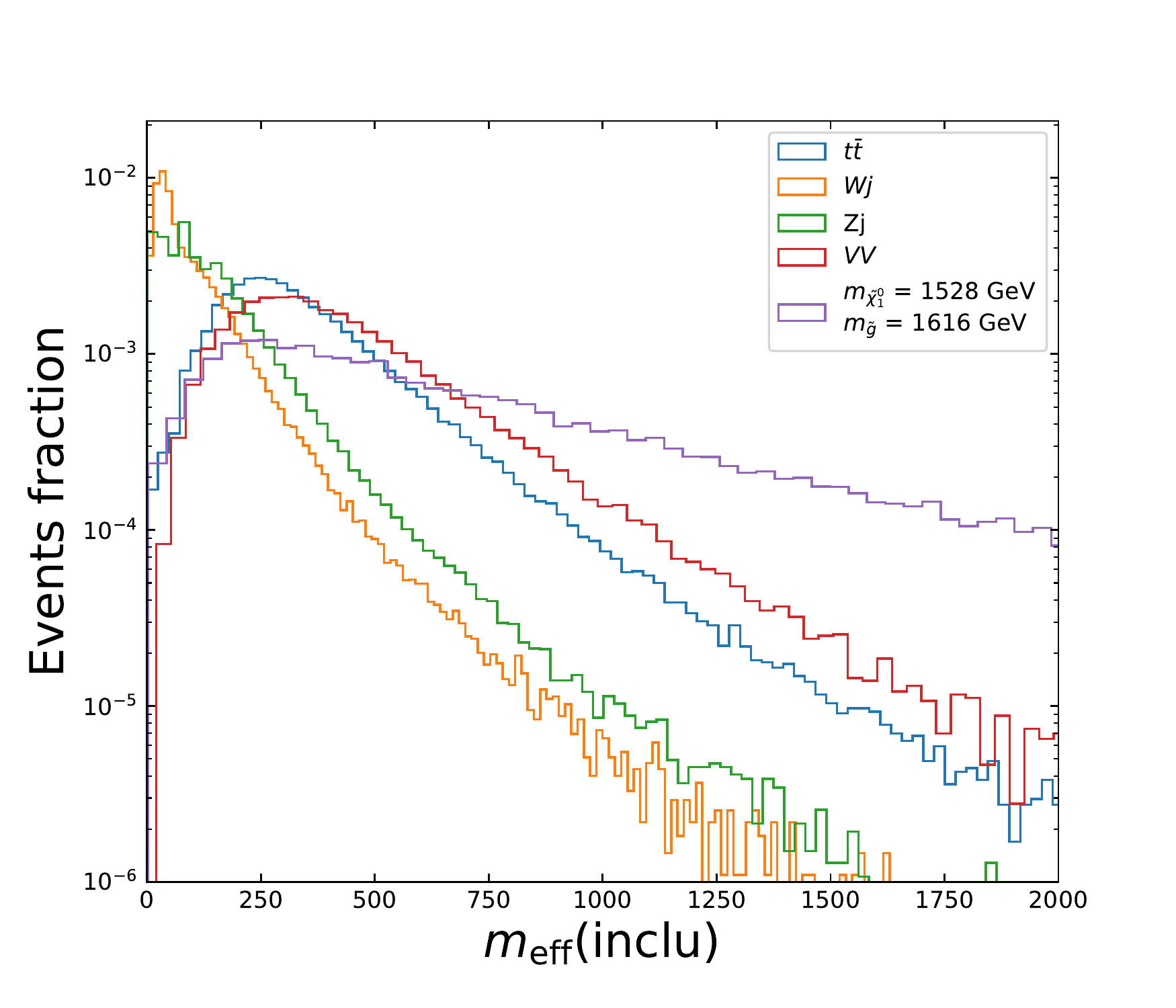}
  \includegraphics[width=7.5cm,height=9cm]{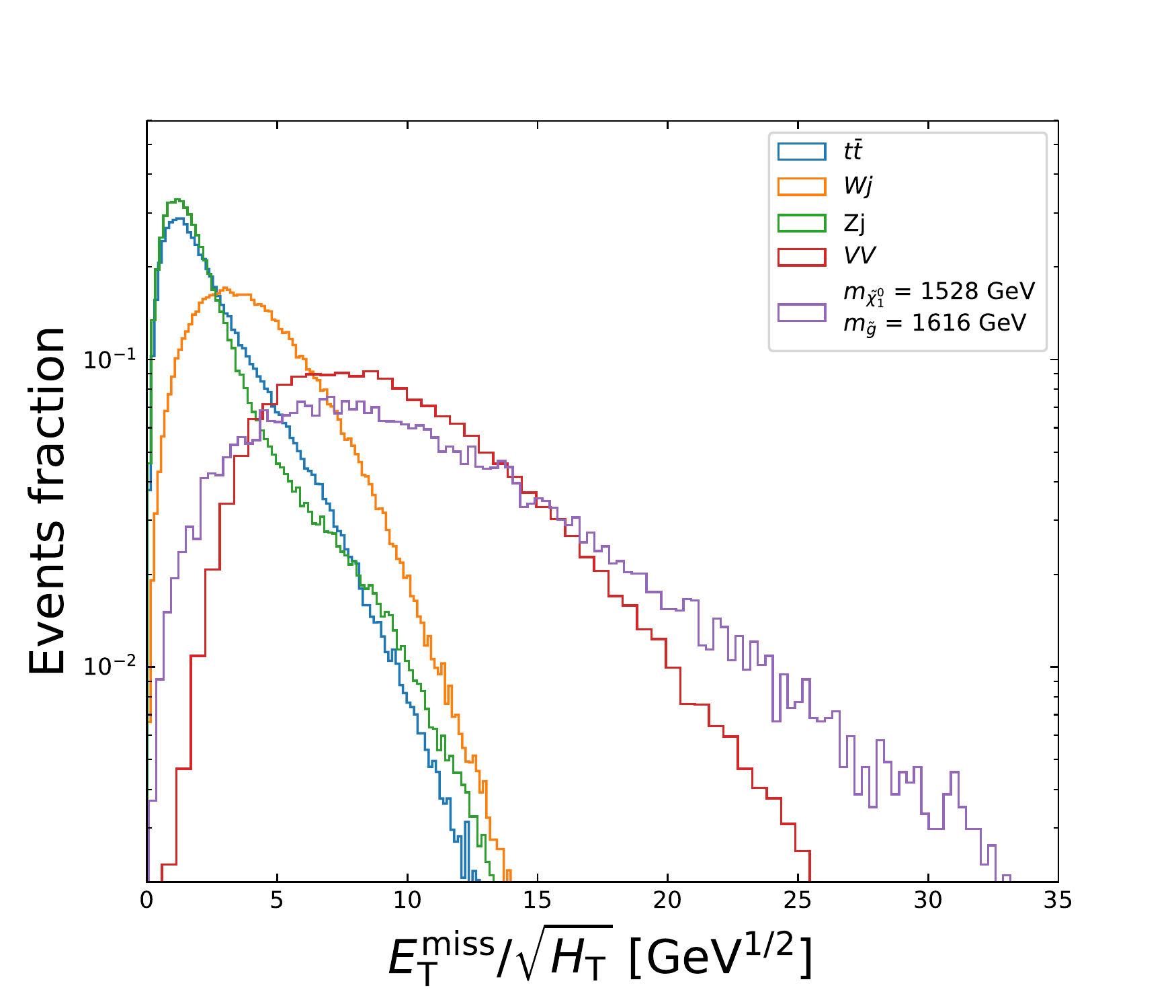}
  \caption{The normalized distributions of $m_{eff}$(inclu) and $E^{miss}_T / \sqrt{H_{T}}$ for the signal and background events at the HE-LHC. The benchmark point is $m_{\tilde g}$ = 1616 GeV and $m_{\tilde \chi_1^0}$ = 1528 GeV.}
  \label{fig:histogram_gluino}
\end{figure}

We firstly check the existing LHC analysis of searching for the 2-6 jets plus missing transverse energy events~\cite{Aaboud:2017vwy} to determine the sensitive signal regions. We find that our gluino coannihilation is sensitive to the signal regions with 2 jets in the final states. Then, we optimize the LHC analysis at the HE-LHC. Due to the small mass difference between the gluino and the neutralino LSP, we will make use of the presence of initial-state radiation (ISR) jets by requiring a higher $p_T$ threshold on the most energetic jet in the event. The signal and background events can be effectively separated by using two kinematical variables, the effective mass $m_{eff}(inclu)$~\cite{Hinchliffe:1996iu} and the ratio of $E^{miss}_T / \sqrt{H_{T}}$, where $m_{eff} (inclu)$ is the scalar sum of transverse momenta of all reconstructed jets with $p_{T} > 50$ GeV, and $H_T$ is the scalar sum of the transverse momenta of all reconstructed jets. From the Fig.~\ref{fig:histogram_gluino}, we can see that most of the signal events lies in the ranges of $m_{eff} (inclu)>700$ GeV. However, the $Zj$ and $Wj$ background events fall off fast, which sequently followed by the $t\bar{t}$ and $VV$ backgrounds. Besides, $m_{eff} (inclu)$ can also strongly suppress the multijet background. Since a hard ISR jet will boost $E_{T}^{miss}$, the signal events predict a larger value of $E_{T}^{miss} / \sqrt{H_{T}}$ than the background events. Such a cut can further enhance the sensitivity of our signal. Therefore, we impose the following event selection criteria:
 \begin{itemize}
   \item The Electrons and muons with $p_{T} > 7$ GeV and $|\eta| < 2.47$ are vetoed.
   \item At least two jets are required, where the leading jet has to satisfy $p_{T}(j_1) > 200$ GeV and other jets should have $p_{T}(j_i) > 50$, where $i>1$.
   \item The events are required to have $E_{T}^{miss} > 250$ GeV.
   \item The azimuthal angular distances between jets and missing enenrgy $\Delta \phi (j_i, p_T^{ miss}) > 0.4, i = 1, 2, (3)$ and $\Delta \phi (j_i, p_T^{ miss}) > 0.2, i > 3$ are required to remove the events with the large $E^{miss}_T$ from the mis-measurement of the jet energy.
   \item In order to cover different kinematical regions, we define six signal regions according to $p_T(j_1)$, $m_{eff}(inclu)$ and $E^{miss}_T / \sqrt{H_{T}}$ in Tab.~\ref{table_gluino}
\end{itemize}

\begin{table}[ht]
  \centering
\begin{tabular}{ccccccc}
  \hline
  SRs & SR1 & SR2 & SR3 & SR4 & SR5 & SR6 \\
  \hline
  $p_\mathrm{T}(j_1)$ [GeV] &250&300&350&400&600&700 \\
  $E_\mathrm{T}^\mathrm{miss}/\sqrt{H_\mathrm{T}} $ [GeV$^{1/2}$] $ >  $ & 16 & 18 & 18 & 18 & 26 & 16\\
  $m_\mathrm{eff}$(inclu) [GeV] $>$ ~~&~~ 1200 ~~ &~~ 1600 ~~ &~~ 2000 ~~&~~ 2400 ~~&~~ 2100~~ &~~ 1300 ~~\\
  \hline
\end{tabular}
  \caption{Six signal regions are defined by the values of $p_T(j_1)$, $m_{eff}(inclu)$ and $E^{miss}_T / \sqrt{H_{T}}$ in the multijet plus missing transverse energy events from the process $pp \to \tilde{g}\tilde{g} +X$ for the gluino coannihilation at the HE-LHC.}
  \label{table_gluino}
\end{table}

\begin{figure}[ht]
  \centering
  \includegraphics[width=15cm,height=9cm]{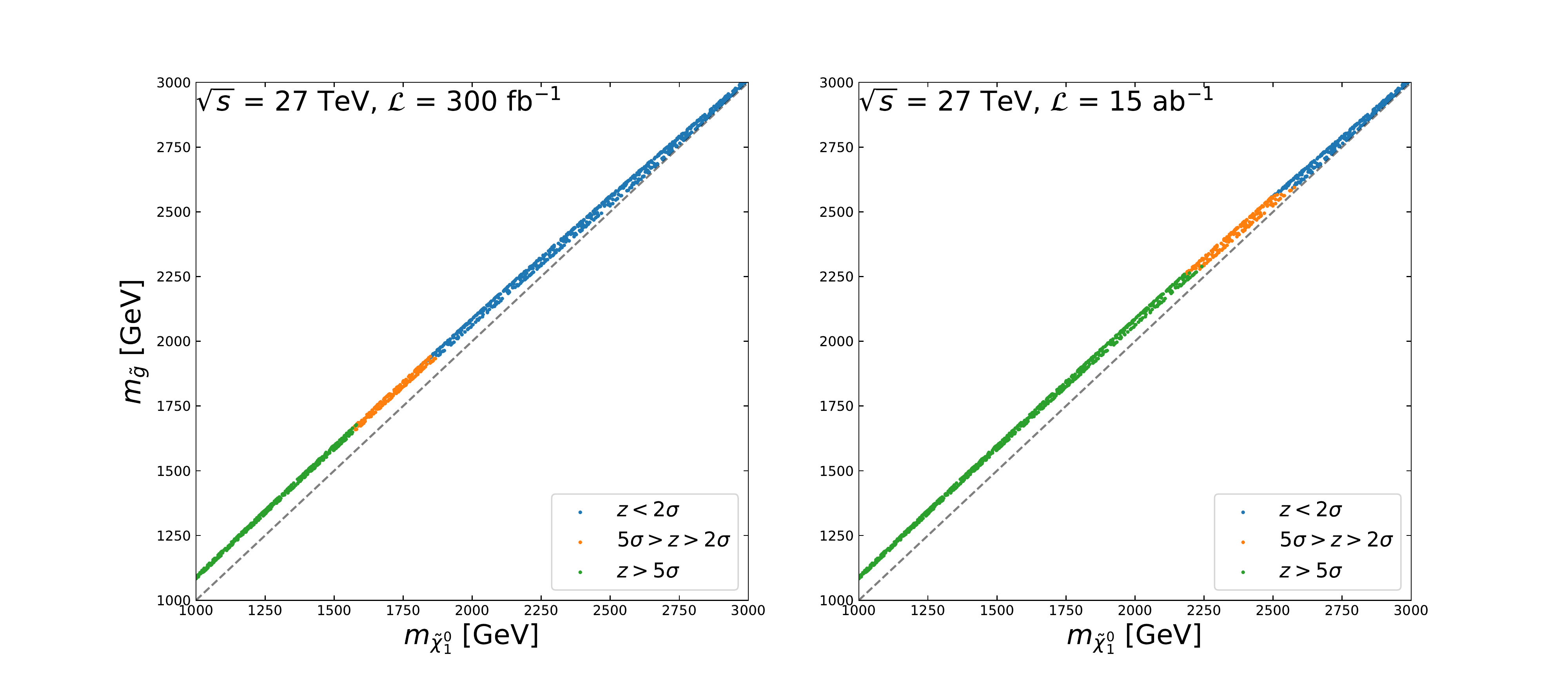}
  \caption{The statistical significance $Z$ of the multijets plus missing transverse energy events from the process $pp \to \tilde{g}\tilde{g} +X$ for the gluino coannihilation at the HE-LHC with the integrated luminosity ${\cal L} = 300$ fb$^{-1}$ and 15 ab$^{-1}$. The projected samples are those satisfying the constraints in Fig.~\ref{fig:mass-range}. The green, orange and blue bullets correspond to $Z>5\sigma$, $2\sigma<Z<5\sigma$ and $Z<2\sigma$, respectively.}
  \label{fig:lhc_gluino}
\end{figure}

In Fig.~\ref{fig:lhc_gluino}, we calculate the signal significance of the multijets plus missing transverse energy events from the process $pp \to \tilde{g}\tilde{g}$ for each surviving sample in the gluino coannihilation at the HE-LHC. From Fig.~\ref{fig:lhc_gluino}, we can see that the significance becomes small with the increase of the gluino mass. The neutralino DM with a mass less than about 1.6 TeV will be probed at $Z \geq 5\sigma$ level at the HE-LHC with the luminosity ${\cal L} = 300$ fb$^{-1}$. Such a mass reach can be enhanced up to about 2.2 TeV if the luminosity increases to 15 ab$^{-1}$. On the other hand, the HE-LHC will be able to exclude the neutralino DM mass $m_{\tilde \chi_1^0}$ in gluino coannihilation up to about 1.9 TeV and 2.6 TeV at $Z=2\sigma$ level, respectively.

\subsection{Stop coannihilation}

\begin{figure}[ht]
  \centering
  \includegraphics[width=12cm,height=8cm]{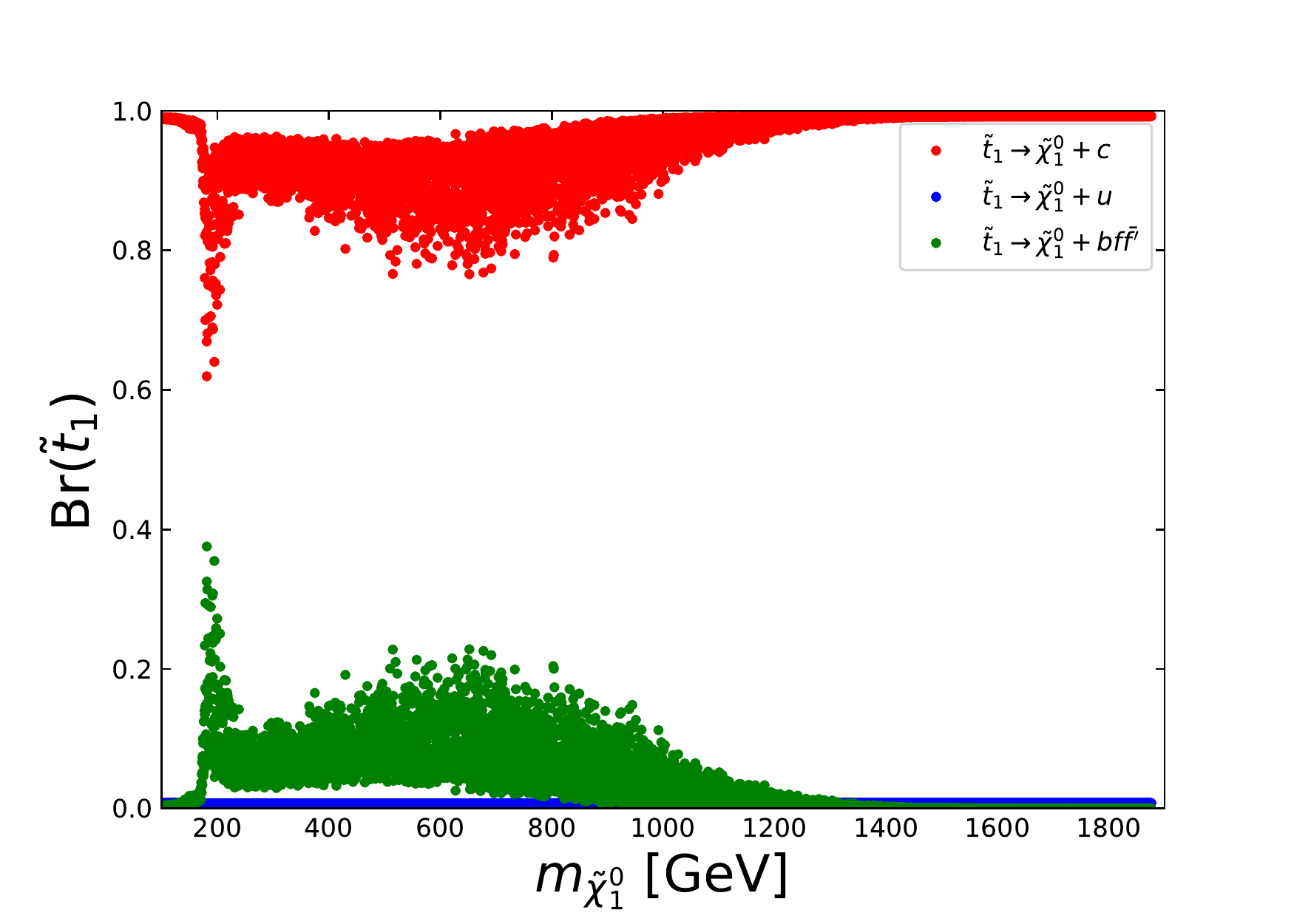}
  \caption{The branching ratio of $\tilde{t}_1$ in the stop coannihilation.}
  \label{fig:br_stop}
\end{figure}
\begin{figure}[ht]
  \centering
  \includegraphics[width=6cm,height=4cm]{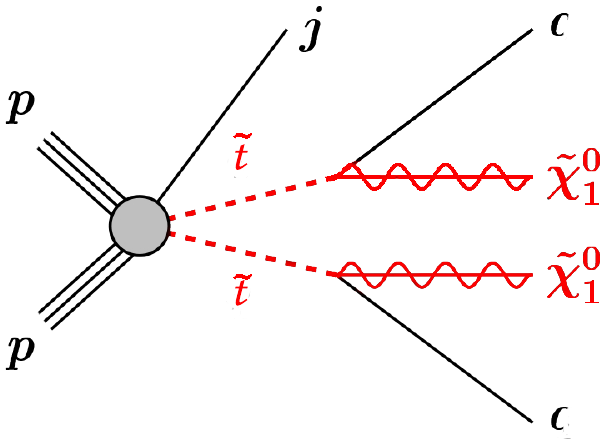}
  \caption{The schematic diagram of the production process $pp \to \tilde{t}_1\tilde{t}^*_1j$ with the sequent decay $\tilde{t}_1 \to c\tilde{\chi}^0_1$ at the HE-LHC.}
  \label{fig:feyn_stop}
\end{figure}

The dominant stop pair production processes are $gg \to \tilde{t}\tilde{t}^*$ and $q\bar{q} \to \tilde{t}\tilde{t}^*$ at the HE-LHC. In the stop coannihilation, From Fig.~\ref{fig:br_stop}, we can see that the stop mainly decays through the loop-induced flavor changing neutral current process $\tilde{t} \to c\tilde \chi_1^0$, which is followed by the three-body decay channel $\tilde{t}_1 \to b f\bar{f}^{'} \tilde{\chi}^0_1$. Besides, the stop co-annihilation requires the stop mass to be close to the LSP neutralino mass so that the light jet from the stop decay is usually too soft to be detected. Thanks to the initial state radiation (ISR) jets, we can boost the stop-pair system to produce the large $E^{miss}_T$ to trigger events, and then suppress backgrounds. Therefore, we utilize the monojet events from the process $pp \to \tilde{t}\tilde{t}^* j+ X$ to probe the stop coannihilation at the HE-LHC, as shown in Fig.~\ref{fig:feyn_stop}. The dominant SM backgrounds come from $Z(\to v\bar v)$ + jets, $W(\to \ell v)$ + jets $(\ell = e, \mu, \tau)$ and $t\bar t$ events. 

\begin{figure}[h]
\includegraphics[width=7.5cm,height=9cm]{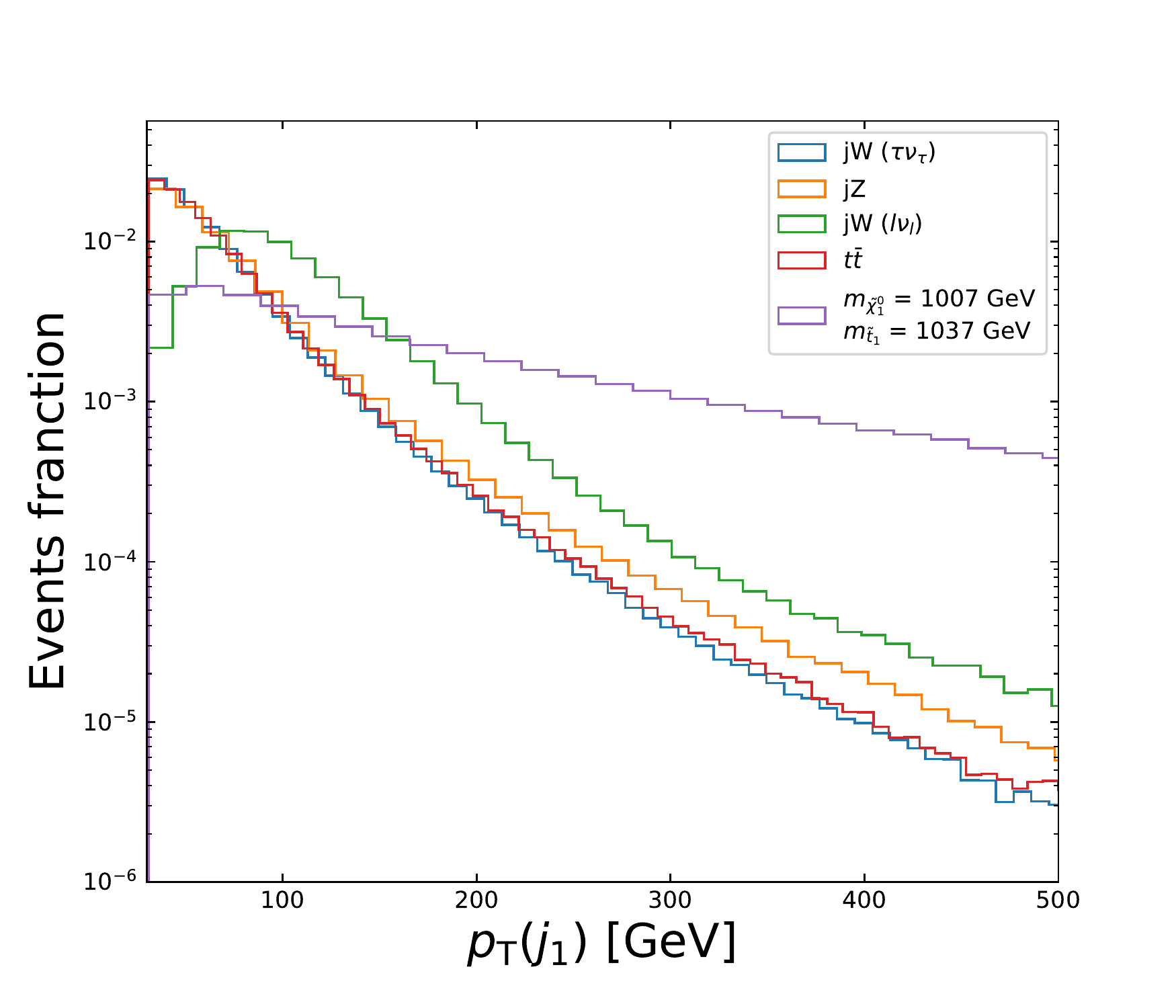}
\includegraphics[width=7.5cm,height=9cm]{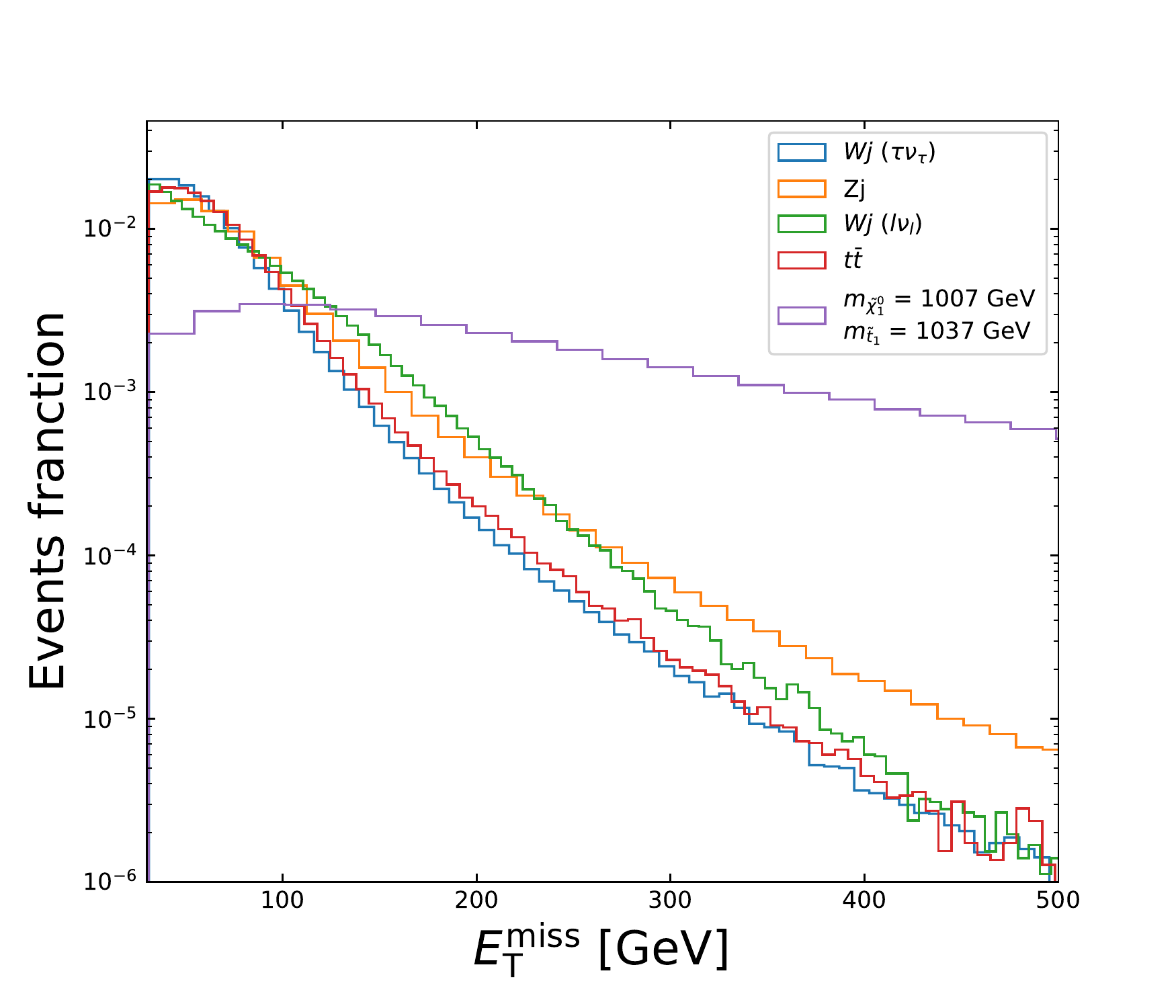}
\caption{The normalized distributions of the leading jet $p_{T}(j_1)$ and $E_{T}^{miss}$ of the signal and background events at the HE-LHC. The benchmark point is $m_{\tilde t_1}$ = 1037 GeV and $m_{\tilde \chi_1^0}$ = 1007 GeV.}
\label{fig:histogram_stop}
\end{figure}

In Fig.~\ref{fig:histogram_stop}, we present the normalized distributions of the leading jet $p_T(j_1)$ and the $E_T^{miss}$ for the signal and backgrounds. It can be seen that the signal has much harder leading jet than the backgrounds. Since the hardness of the event is determined by the $p_T$ of the leading jet, the slopes of the leading jet $p_T$ distribution in high $p_T$ region are almost the same for different stop mass. The $E_T^{miss}$ distribution of the signal events has the slowest fall-off. We expect that a hard cut on $E_T^{miss}$ will remove the backgrounds effectively. Therefore, we impose the following event selection criteria:
 \begin{itemize}
   \item Events are selected with a leading jet with $p_T (j_1) > 200$~GeV and $|\eta| < 2.4$.
   \item Events having muons with $p_T > 10$~GeV or electrons with $p_T > 20$~GeV in the final states are vetoed.
   \item At most four jets with $p_T > 30$~GeV and $|\eta| < 2.8$ are allowed.
   \item The azimuthal angle between the leading jet and missing transverse momentum $\Delta \phi (j_1, p_T^{ miss}) > 0.4 $ is required to remove the events with the large $E^{miss}_T$ from the mis-measurement of the jet energy.
   \item Five signal regions, $E_T^{miss}>$ 200, 400, 600, 800, 1000 GeV, are defined in our analysis.
\end{itemize}

\begin{figure}[th]
\centering
	\includegraphics[width=15cm,height=9cm]{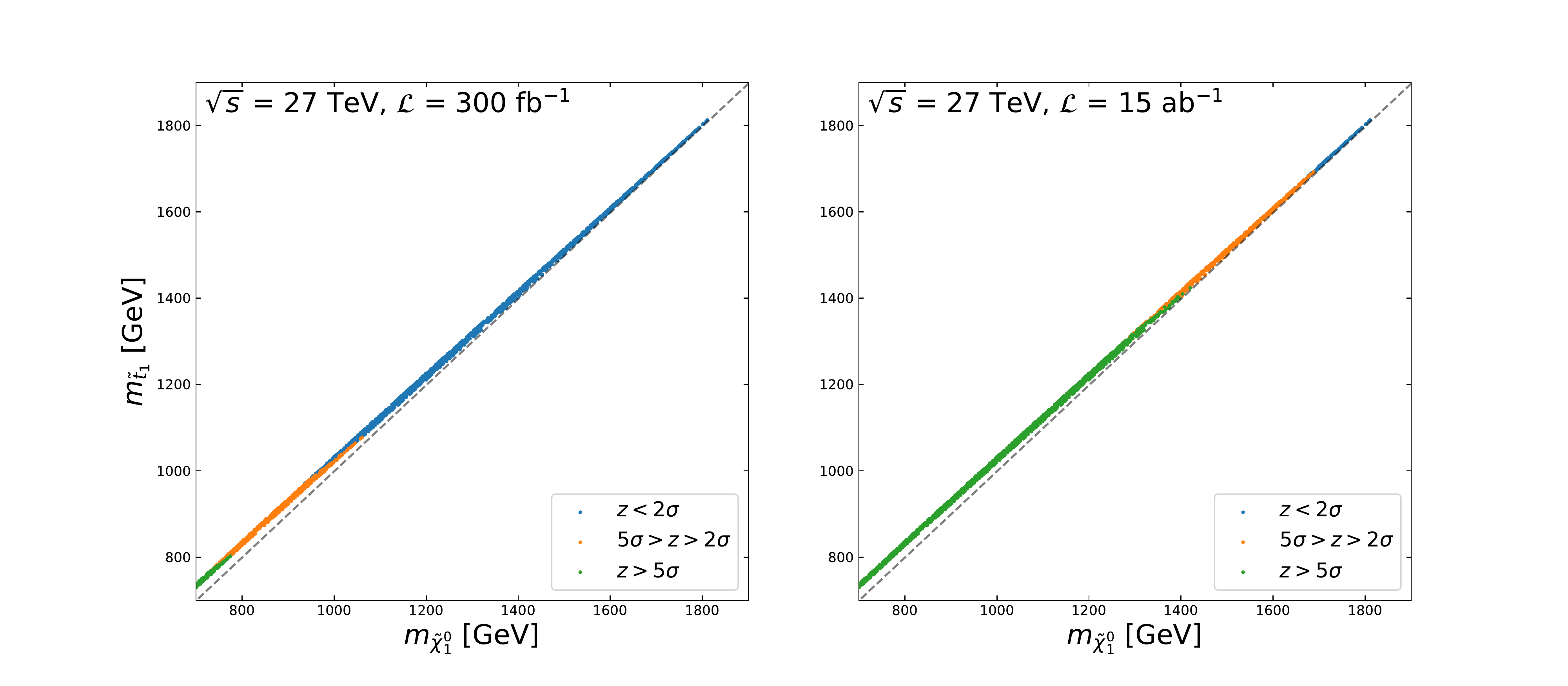}
	\caption{Same as Fig.~\ref{fig:lhc_gluino}, but for the monojet events from the process $pp \to \tilde{t}_1 \tilde{t}^*_1 j \to E^{miss}_T$+jets. }
	\label{fig:lhc_stop}
\end{figure}

In Fig.~\ref{fig:lhc_stop}, we project the surviving samples for stop coannihilation on the plane of $m_{\tilde{t}_1}$ versus $m_{\tilde{\chi}^0_1}$ at the HE-LHC. It can be seen that the HE-LHC is able to probe the DM with a mass $m_{\tilde \chi_1^0}<$ 800 and 1400 GeV at $Z \geq 5\sigma$ level for the luminosity ${\cal L}=300$ fb$^{-1}$ and 15 ab$^{-1}$, respectively. On the other hand, if there was no significant excess, we point out that the DM mass $m_{\tilde \chi_1^0}$ will be excluded up to 1.1 TeV and 1.7 TeV at $Z=2\sigma$ level, respectively.

\subsection{Wino coannihilation}

\begin{figure}[ht]
  \centering
  \includegraphics[width=12cm,height=8cm]{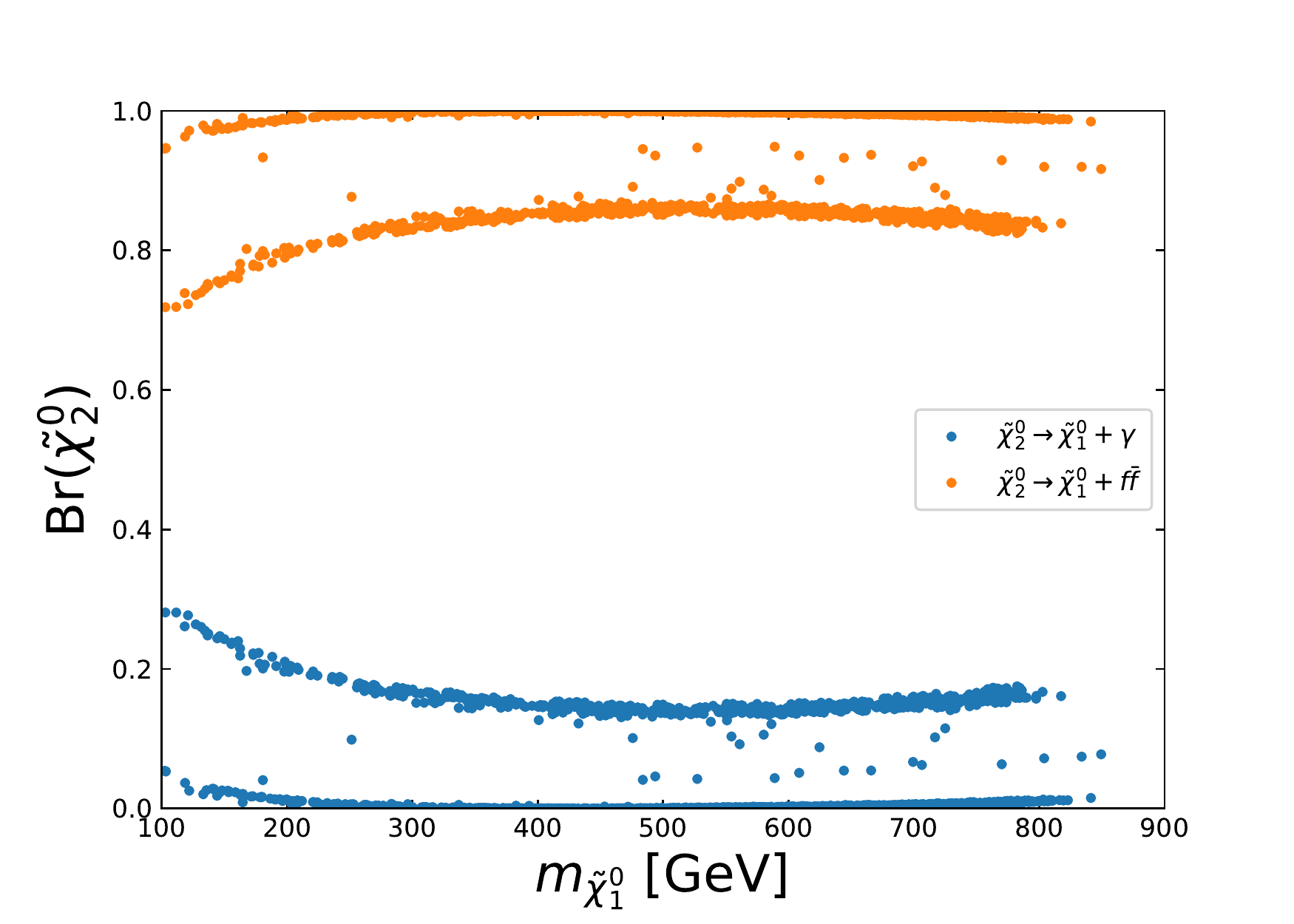}
  \caption{The branching ratio of $\tilde{\chi}^\pm_1$ and $\tilde{\chi}^0_2$ in the wino coannihilation.}
  \label{fig:br_wino}
\end{figure}

\begin{figure}[ht]
  \centering
  \includegraphics[width=6cm,height=4cm]{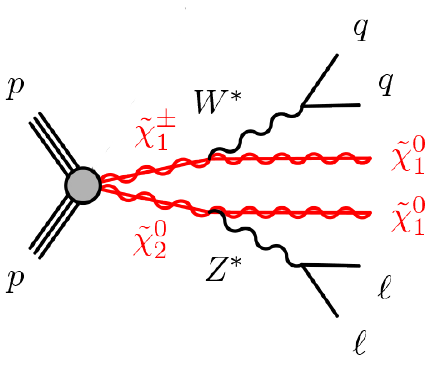}
  \caption{The schematic diagram of the production process $pp \to \tilde{\chi}^\pm_1\tilde{\chi}^0_2$ with the sequent decays $\tilde{\chi}^0_2 \to Z(\to \ell^+\ell^-)\tilde{\chi}^0_1$ and $ \tilde{\chi}^\pm_1 \to W(\to q\bar{q})\tilde{\chi}^0_1$.}
  \label{fig:feyn_wino}
\end{figure}

The wino-like electroweakinos $\tilde{\chi}^\pm_1$ and $\tilde{\chi}^0_2$ can be produced through the Drell-Yan process $pp \to \tilde{\chi}^\pm_1\tilde{\chi}^0_2$ at the HE-LHC. From Fig.~\ref{fig:br_wino}, we can see that the next-to-lightest neutralino $\tilde{\chi}^0_2$ will dominantly decay via the three-body process $\tilde \chi_2^0 \to \tilde \chi_1^0 Z^* \to \tilde \chi_1^0 f \bar f$. Subsequently, the loop process $\tilde{\chi}^0_2 \to \gamma \tilde{\chi}^0_1$ can have the branching ratio of $10\% \div 30\%$, which may provide a distinctive signature of a soft photon plus large missing transverse energy with a hard ISR jet at the LHC~\cite{Han:2014xoa}. The lightest chargino $\tilde \chi_1^\pm$ will proceed through the three-body decay process $\tilde \chi_1^\pm \to \tilde \chi_1^0 {W^\pm}^* \to \tilde \chi_1^0 f \bar f^\prime $. Given the compressed spectrum of wino coannihilation, we perform the Monte Carlo simulation of the process $pp \to \tilde \chi_1^\pm \tilde \chi_2^0 \to \ell^+\ell^- + E^{miss}_T$+ jets at the HE-LHC, as shown in Fig.~\ref{fig:feyn_wino}. The dominant SM backgrounds in this scenario are Drell-Yan process $pp \to \gamma^* / Z^*$+ jets, the dibosons ($WW,\ WZ,\ ZZ$) and the single top. In contrast with the conventional monojet analysis for the compressed electroweakinos~\cite{Han:2013usa}, there is no upper limit on the number of jets in our analysis. Instead, we require a small transverse mass $m_T(\ell,\nu_\ell)$ to suppress the $t\bar{t}$ background. Besides, the soft lepton pair from the decay of $\tilde{\chi}^0_2$ can be used as a handle to reduce the huge background $V$+jets.

\begin{figure}[ht]
  \centering
  \includegraphics[width=7.5cm,height=9cm]{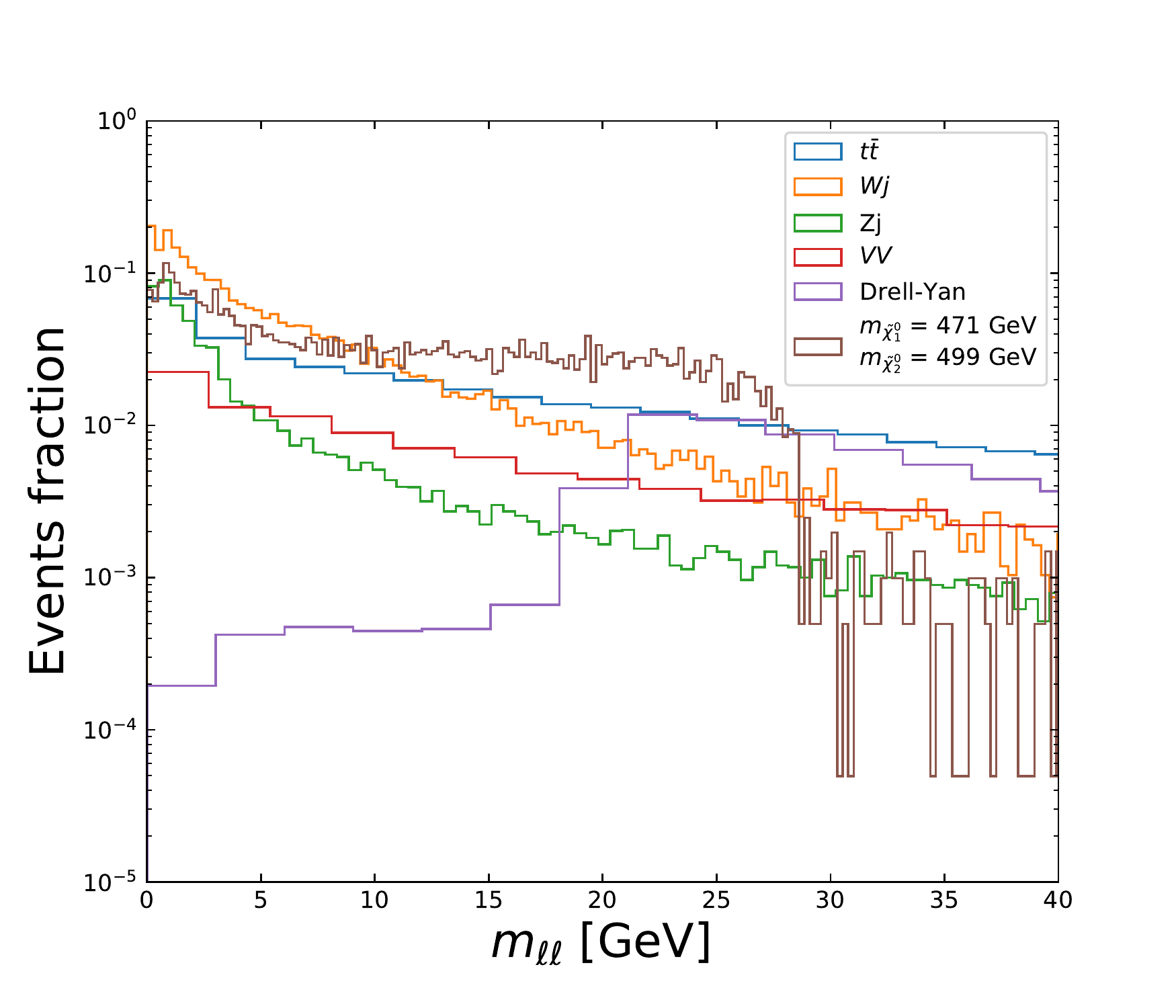}
  \includegraphics[width=7.5cm,height=9cm]{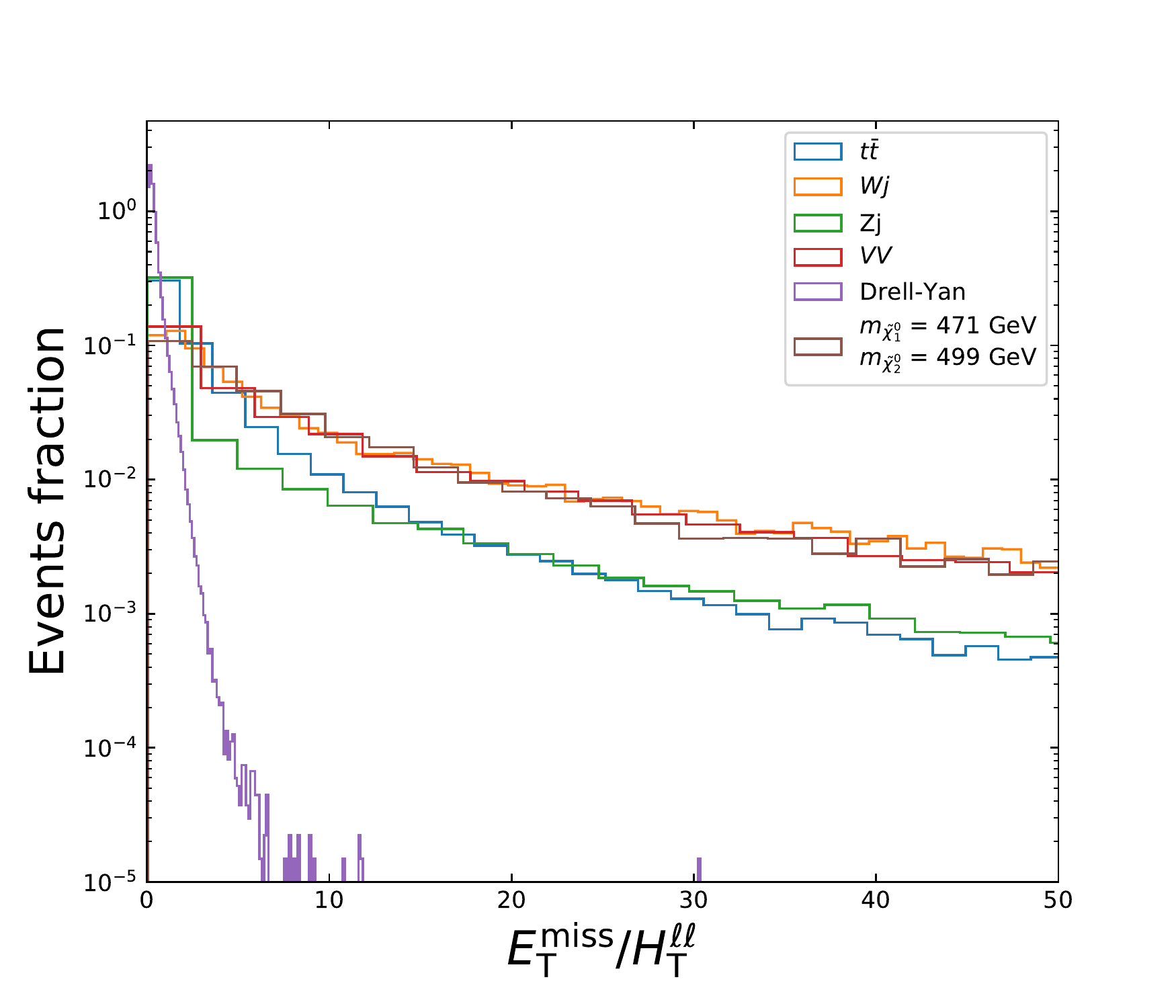}
  \caption{The normalized distributions of $m_{\ell \ell}$ and $E_{T}^{miss} / H_{T}^{lep}$ of the signal and background events at the HE-LHC. The benchmark point is $m_{\tilde \chi_2^0}=499$ GeV, $m_{\tilde \chi_1^\pm}=498$ GeV and $m_{\tilde \chi_1^0}=471$ GeV.}
  \label{fig:histogram_wino}
\end{figure}
In Fig.~\ref{fig:histogram_wino}, we show the normalized distribution of the dilepton invariant mass $m_{\ell \ell}$ and the ratio of $E_{T}^{miss} / H_{T}^{lep}$ of the signal and background events. We can find that the signal has more events in the range of the invariant mass $m_{\ell \ell} \leq 30$ GeV~\cite{Baer:2014kya,Barr:2015eva}, which can highly suppress the $\gamma^* / Z^{(*)} (\to \ell \ell)+ jets$ and $t\bar{t}$ backgrounds. Besides, it can also hurt the fake/non-prompt leptons effectively~\cite{Han:2014kaa}. On the other hand, the signal events of the compressed electroweakinos production predict a small value of the scalar sum of the lepton transverse momenta $H_{T}^{lep} = p_{T}^{\ell_1} +  p_{T}^{\ell_2}$. Thus, the ratio $E_{T}^{miss} / H_{T}^{lep}$ can significantly reduce the Drell-Yan and QCD multijet backgrounds~\cite{Sirunyan:2018iwl}.

\begin{figure}[ht]
  \centering
  \includegraphics[width=15cm,height=9cm]{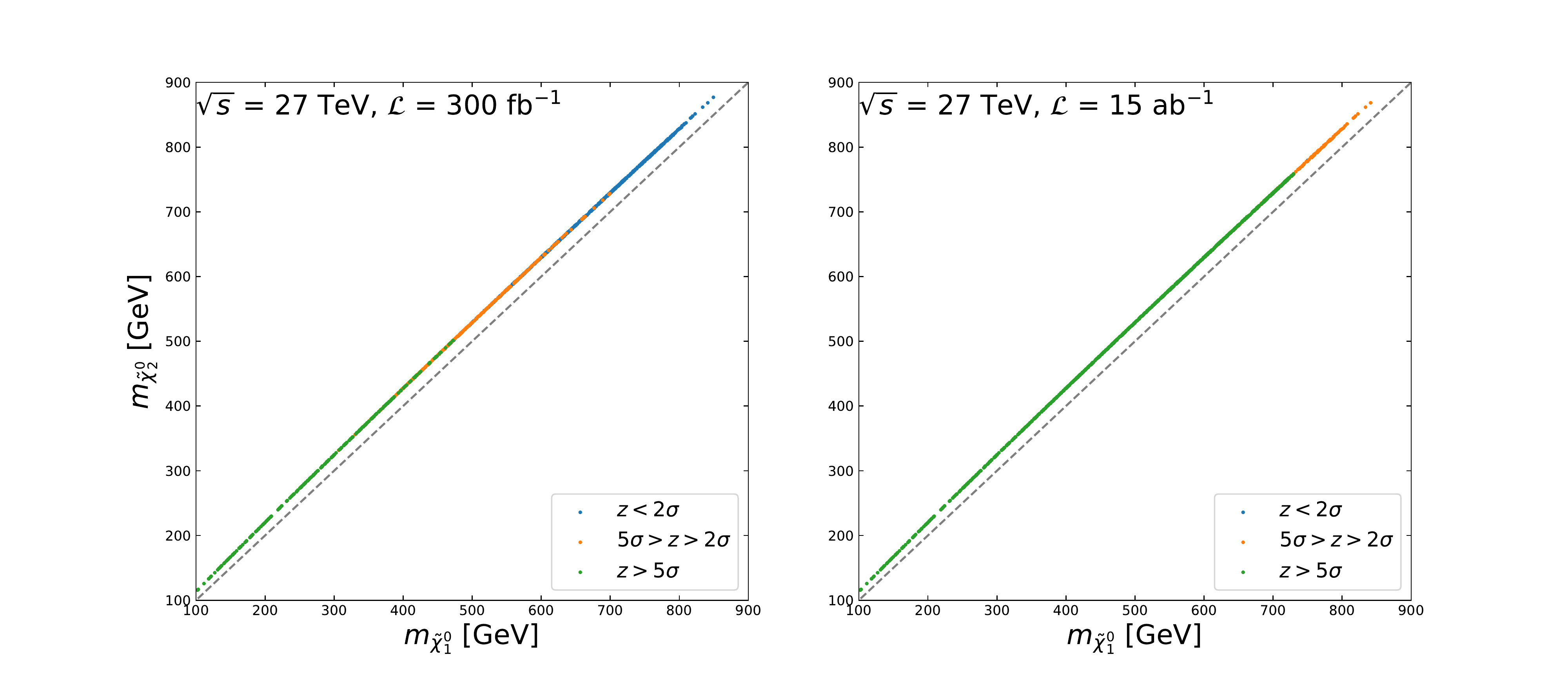}
  \caption{Same as Fig.~\ref{fig:lhc_gluino}, but for the soft lepton pair plus missing transverse energy events from the process $pp \to \tilde{\chi}^\pm_1\tilde{\chi}^0_2 \to \ell^+\ell^- + E^{miss}_T$+ jets.}
  \label{fig:lhc_wino}
\end{figure}

According to the features of above distributions, we perform the following kinematical cuts.
\begin{itemize}
  \item Events are required to have exactly two same flavor opposite sign leptons ($e^+e^-$ or $\mu^+\mu^-$). The leading and subleading leptons should have transverse momentum larger than 5 and 4 GeV, respectively. The separation $\Delta R=\sqrt{\Delta\eta^2+\Delta\phi^2}$ between two leptons are between 0.05 and 2.
  \item We select events with $E_{T}^{miss} > 200$ GeV and require the transverse momentum of the leading jet $p_T(j_1)>$ 100 GeV and $\Delta \phi (j_1, E_{T}^{miss}) > 2$. We veto any events with $b$-jets.
  \item To reduce the mismeasurement effect, the events with the azimuthal angle between any jet and $E_{T}^{miss}$ smaller than 0.35 is discarded.
  \item We require the transverse mass $m_{T}(\ell_1, E_{T}^{miss})$ of the leading lepton to be less than 60 GeV, which can reduce $t\bar{t}$ significantly.
  \item We define ten signal regions by the values of $m_{\ell \ell}$ and $E_{T}^{miss} / H_{T}^{lep}$, as shown Table~\ref{table_wino}.
\end{itemize}

\begin{table}
  \centering
\begin{tabular}{cccccc}
  \hline
  SRs & SR1 & SR2 & SR3 & SR4 & SR5  \\
  \hline
  $m_{\ell_1 \ell_2}$[GeV] $\in$ & [5, 10] & [10, 15] & [15, 20] & [20, 25] & [25,30]  \\
  $E_{\rm T}^{\rm miss}/H_{\rm T}^{\rm lep} >$ & 12 & 10 & 8 & 6 & 6  \\
  \hline\hline
  SRs  & SR6 & SR7 & SR8 & SR9 & SR10 \\
  $m_{\ell_1 \ell_2}$[GeV] $\in$ & [1, 3] $\cup$ [5, 30] & [5, 30] & [10, 30] & [15, 30] & [20, 30]\\
  $E_{\rm T}^{\rm miss}/H_{\rm T}^{\rm lep} >$ & 14 & 12 & 10 & 8 & 6 \\
  \hline
\end{tabular}
  \caption{Ten signal regions are defined by the values of $m_{\ell \ell}$ and $E_{T}^{miss} / H_{T}^{lep}$.}
  \label{table_wino}
\end{table}

In Fig.~\ref{fig:lhc_wino}, we present the signal significance of the soft lepton pair plus missing transverse energy events from the process $pp \to \tilde{\chi}^\pm_1\tilde{\chi}^0_2+X$ at the HE-LHC. Compared with the Fig.~\ref{fig:mass-range}, we notice that the $2\sigma$ exclusion limit of the neutralino DM mass will be extended from about 180 GeV at the current LHC to 560 GeV at the HE-LHC with the luminosity ${\cal L}=300$ fb$^{-1}$. Besides, the neutralino DM with the mass less than about 470 GeV can be probed at $Z \geq 5\sigma$ level at the HE-LHC, which will be improved up to about 620 GeV as the luminosity increasing to $15$ ab$^{-1}$.

\subsection{Stau coannihilation}

\begin{figure}[ht]
  \centering
  \includegraphics[width=6cm,height=4cm]{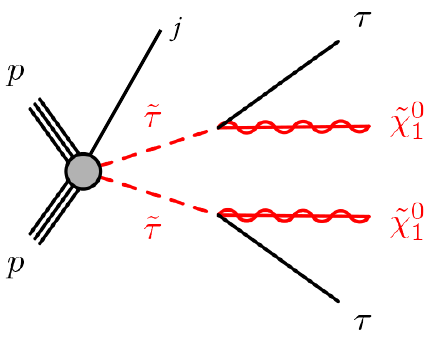}
  \caption{The schematic diagram of the production process $pp \to \tilde{\tau} \tilde{\tau}^* j + X$ with the sequent decays $\tilde{\tau} \to \tau \tilde{\chi}^0_1$.}
  \label{fig:feyn_stau}
\end{figure}
The search for a light stau is experimentally difficult due to its extremely low production rate at the LHC. The staus can be produced directly in pairs through the channel $pp \to \tilde{\tau}\tilde{\tau}^*$. Then, the stau decays with a branching fraction of 100\% to the SM tau-lepton and the LSP neutralino. This will give the signature of $\tau^+\tau^- + E^{miss}_T$ at the HE-LHC. The hadronic decay of $\tau$ lepton has the largest branching fraction and thus final states with a $\tau_h$ provide the best experimental sensitivity. Signal events would thus be characterised by the presence of two sets of close-by hadrons and large $E^{miss}_T$ originating from the invisible LSP and neutrinos. Events are further categorized into regions with different $E^{miss}_T$, to examine different stau mass scenarios. During LHC Run-1, only a narrow parameter region around a stau mass of 109 GeV and a massless lightest neutralino could be excluded by the LHC experiments. Such a mass limit has been extended to 300$\div$400 GeV in the ATLAS Run-2 search of stau~\cite{ATLAS:2019ucg}.

In the stau coannihilation, the small mass difference between stau and LSP neutralino results in low $p_T$ visible decay products, making it difficult to identify $\tau$ lepton. Besides, the semi-leptonic decays of $\tau$ lepton leads to lower average $p_T$ than hadronic decays, while also being largely indistinguishable from prompt production of electrons and muons. Therefore, we will study the events with one soft hadronically decaying tau lepton and missing transverse energy recoiling against a hard $p_T$ jet from ISR, as shown in Fig.~\ref{fig:feyn_stau}. The dominant SM backgrounds include the Drell-Yan+jets, $V$+jets, $t\bar{t}$ and the dibosons ($WW$, $WZ$, $ZZ$).

\begin{figure}[ht]
  \centering
  \includegraphics[width=7.5cm,height=8cm]{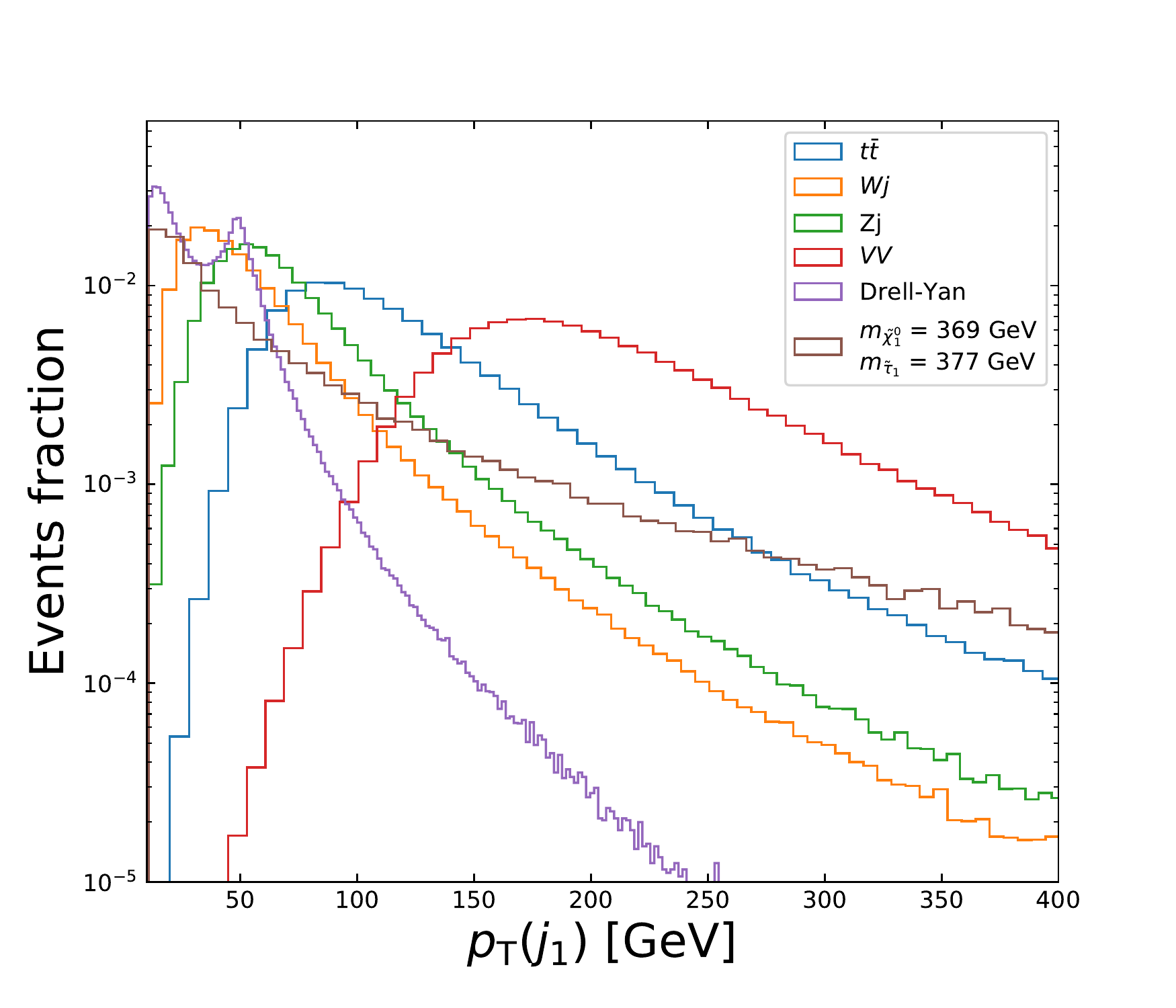}
  \includegraphics[width=7.5cm,height=8cm]{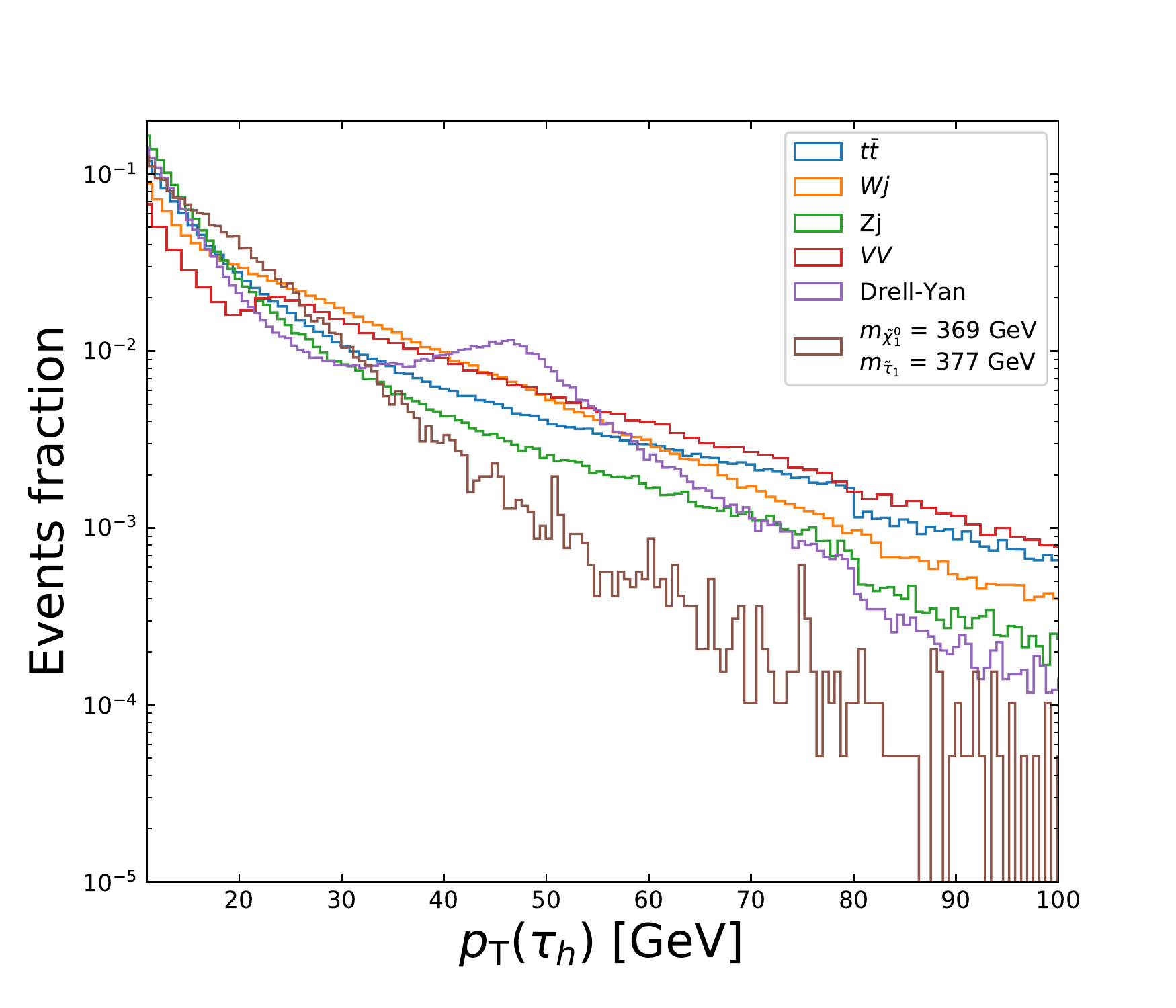}
  \includegraphics[width=7.5cm,height=8cm]{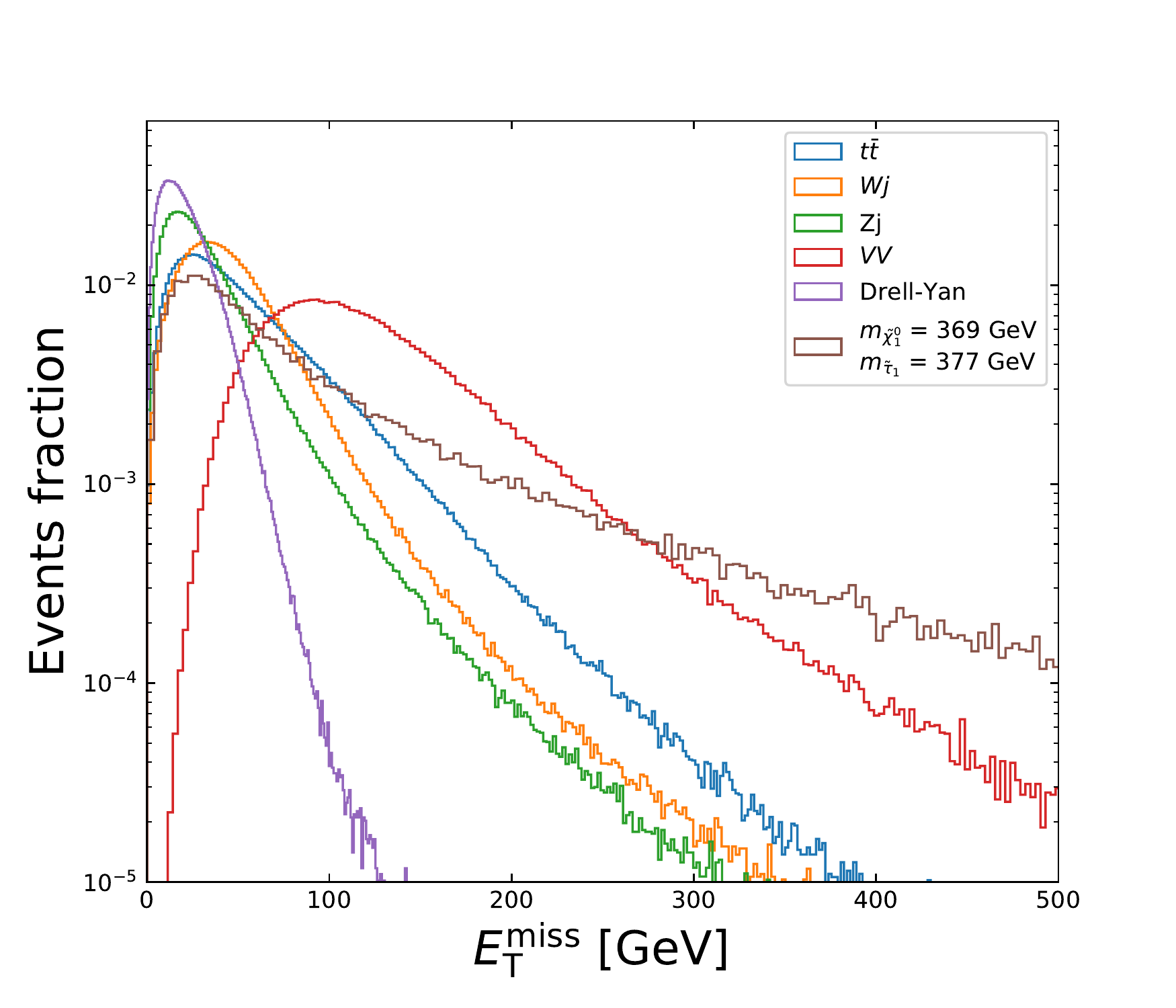}
  \includegraphics[width=7.5cm,height=8cm]{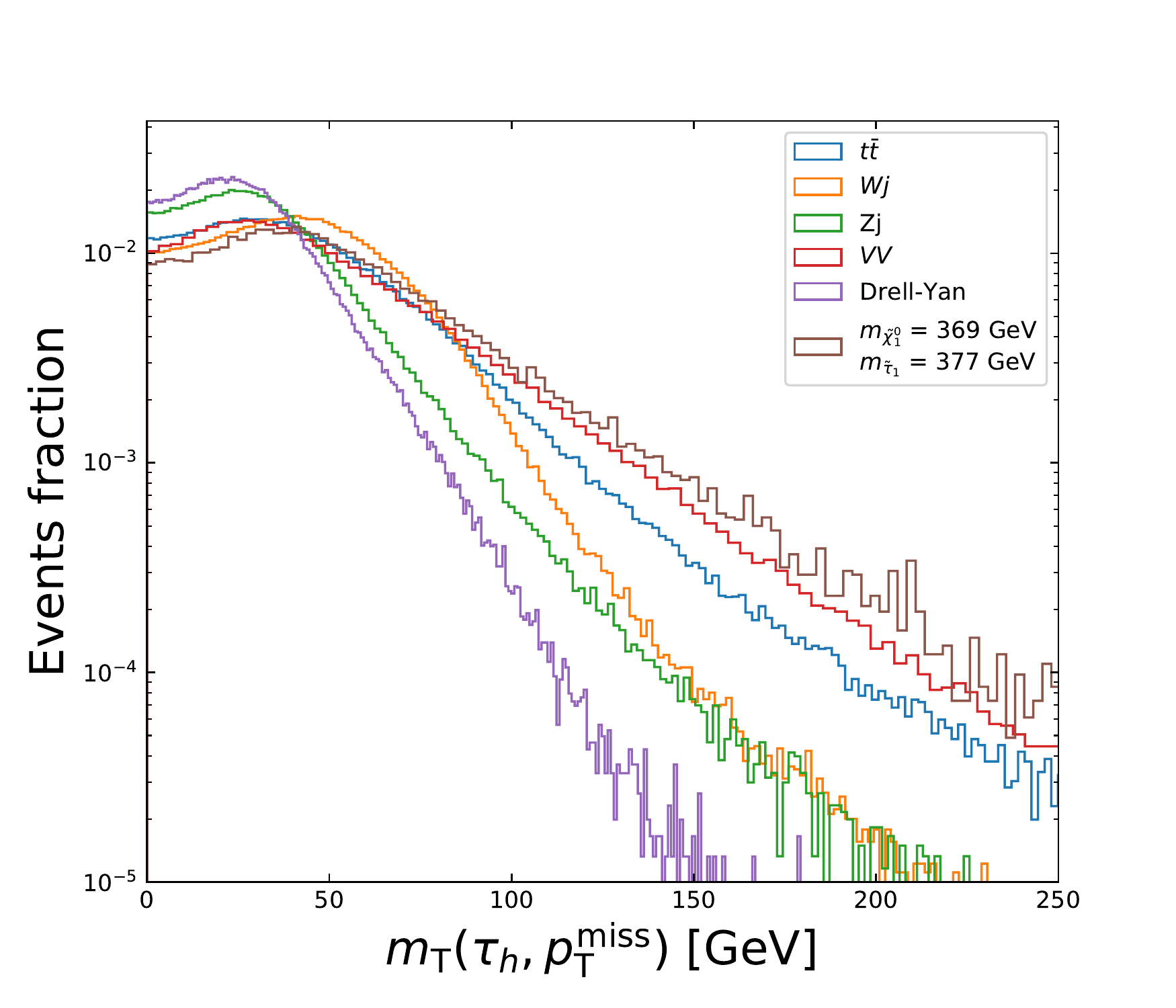}
  \caption{The normalized distributions of $p_T(j_1)$, $p_T(\tau_h)$, $E^{miss}_T$ and $m_{T_{\tau_h}}$ of the signal and backgrounds at 27 TeV HE-LHC. The benchmark point is $m_{\tilde \chi_1^0}$ =369 GeV,  $m_{\tilde \tau_1}$ = 377 GeV.}
  \label{fig:histogram_stau}
\end{figure}

In Fig.~\ref{fig:histogram_stau}, we show the normalized kinematical distributions of the signal and background events. It can be seen that the transverse momentum of the leading jet $p_T(j_1)$ in the signal are much harder than those in the Drell-Yan and $V$+jets backgrounds. On the other hand, the transverse momentum of the hadronic tau $p_T(\tau_h)$ from the decay of stau are very soft, most of which distribute in the region of $p_T(\tau_h)<35$ GeV. In addition, the signal has a larger missing transverse energy $E^{miss}_T$ than all backgrounds. Another sensitive discriminator is the transverse mass $m_{T_{\tau_h}}$ between $\tau_h$ and $p^{miss}_T$, which is defined as
\begin{equation}
m_{T_{\tau_h}}=\sqrt{2p^{miss}_T p_T(\tau_h)(1-\cos\Delta\phi(\tau_h,p^{miss}_T))}.
\end{equation}
As comparison with the background, the signal has more events in the range of high values of $m_{T_{\tau_h}}$ than the backgrounds. For example, when $m_{T_{\tau_h}}>150$ GeV, almost events of the Drell-Yan process will be removed.

In the selection of the signal events, we impose the following criteria to suppress the backgrounds:
\begin{itemize}
  \item Events containing any isolated electron or muon, with $p_{T} > 20$ GeV, are vetoed.
  \item Events with any $b$-jet are rejected. The leading jet $p_T(j_1)$ has to be larger than 100 GeV.
  \item Events are required to have exactly one $\tau_{h}$ with $15 < p_{T}(\tau_{h}) < 35$ GeV and $|\eta(\tau_{h})| < 2.3$. The efficiency of tau tagging is assumed as 60\%.
  \item Jets and $\tau_{h}$ are required to be well separated by a cut of $\Delta R (\tau_{h}, jet) >0.4$, which will reject jets from QCD processes that can mimic the signature of a $\tau_{h}$.
  \item Events with the missing transverse energy $E^{miss}_T$ larger than 230 GeV are required. Then, the multijets events will becomes negligible.
  \item We define three signal regions according to the values of $m_{T_{\tau_h}}$, which is shown in the Tab.~\ref{table_stau}.
\end{itemize}

\begin{table}[ht]
  \centering
\begin{tabular}{cccccccc}
  \hline
  SRs & ~~ SR1 ~~ &~~ SR2 ~~&~~ SR3  \\
  \hline
  $m_{T_{\tau_h}}$[GeV] & $>150$ & $>200$ & $>250$\\
  \hline
\end{tabular}
    \caption{Three signal regions are defined by the values of $m_{T_{\tau_h}}$.}
  \label{table_stau}
\end{table}

\begin{figure}[ht]
  \centering
  \includegraphics[width=15cm,height=9cm]{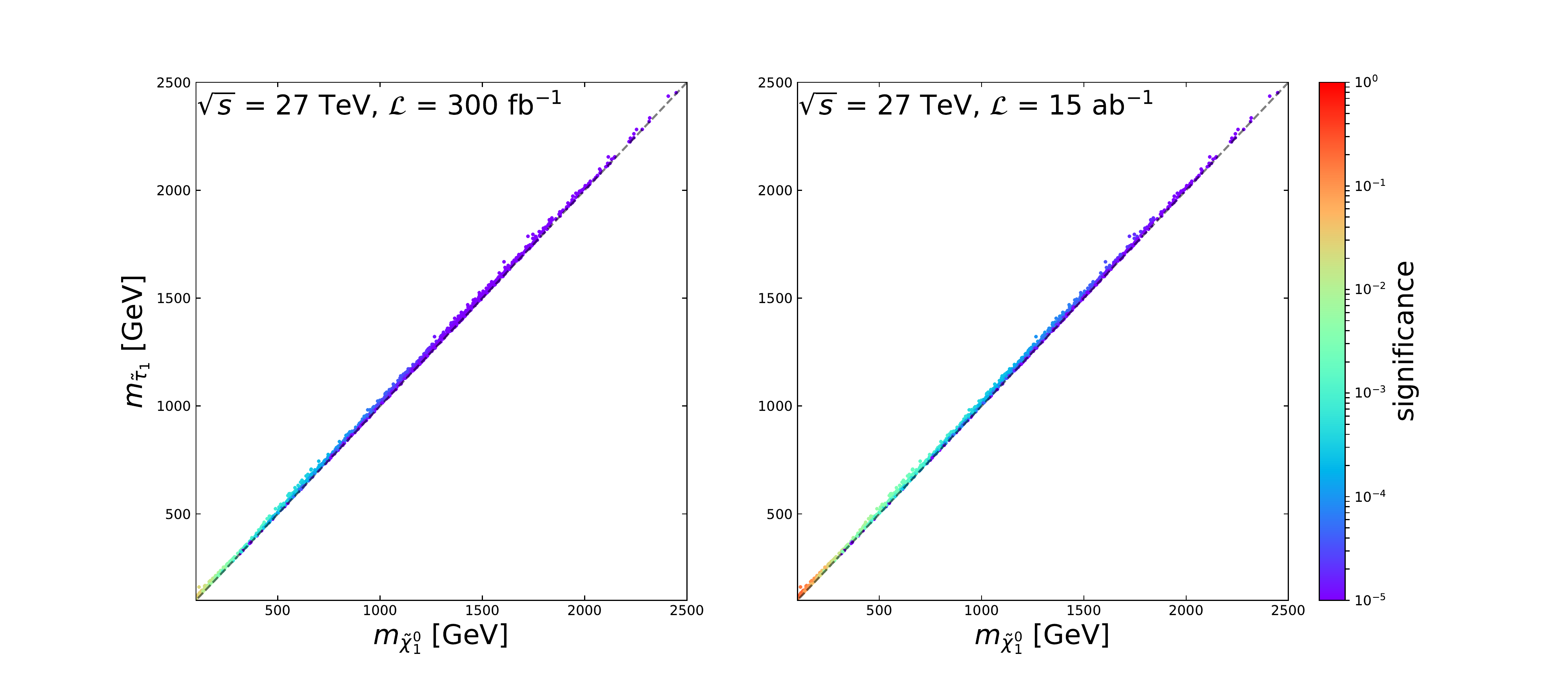}
  \caption{The statistical significance of the process $pp \to j\tilde{\tau}_1 \tilde{\tau}^*_1 \to j+\tau_h+E^{miss}_{T}$ at the HE-LHC.}
  \label{fig:lhc_stau}
\end{figure}
Figure \ref{fig:lhc_stau} shows the statistical significance of the process $pp \to j\tilde{\tau}_1 \tilde{\tau}^*_1 \to j+\tau_h+E^{miss}_{T}$ at the HE-LHC. We find the significance of all samples in the stau coannihilation is less than $2\sigma$. In contrast with the above search for the compressed winos, there are two main reasons for such a poor sensitivity: one is that the cross section of stau pair production is relative small, which is about $1/4$ of the cross section of wino pair production for the same mass. The other is that tau tagging efficiency for the soft tau from stau decay is badly reduced. We also used the proposed analysis with two tagging hadronic tau lepton~\cite{Aboubrahim:2017aen} and found that it reach a similar sensitivity as ours. Besides, the vector boson fusion topologies with taus in the final states have been proposed~\cite{Dutta:2012xe}, however, which has a smaller cross section of the process $pp \to jj\tilde{\tau}\tilde{\tau}^*$ at the HE-LHC.

In above discussions, we present the mass reaches of the gluino, stop, wino and stau in the coannihilations at the HE-LHC. It should be mentioned that the statistical significance will get degraded when systematic uncertainties are taken into account. The determination of the systematic uncertainties due to the high pile-up conditions of the future hadron-collider runs is beyond the scope of this paper. It must be revisited with the real performance of the upgraded ATLAS and CMS detectors. Besides, the machine learning methods have been recently proposed to enhance the sensitivity in the search of sparticles at the LHC~\cite{Albertsson:2018maf,Abdughani:2019wuv,Abdughani:2018wrw,Ren:2017ymm,Caron:2016hib}. We expect that our results may be improved by using those advanced analysis approaches at the HE-LHC.

\section{Conclusions} \label{section4}
In this paper, we investigate the potential of the discovery of the neutralino DM in the gluino, stop, wino, stau coannihilations at the HE-LHC. We carried out our study in the simplified MSSM model that only includes the relevant sparticles in each scenario. We firstly impose the relic density constraint and determine the allowed parameter space of the gluino, stop, wino and stau coannihilations. Since the mass difference between the neutralino DM and its coannihilating partner is usually small, the discovery of coannihilating partner will also provide a measurement of the neutralino DM mass. Thus, we perform the Monte Carlo simulations to investigate the observability of gluino, stop, wino and stau in each coannihilation scenario at the HE-LHC. Our analysis strategies include the multijet with $E^{miss}_T$, the monojet, the soft lepton pair plus $E^{miss}_T$, and the monojet plus a hadronic tau. In this end, it is found that the neutralino DM mass can be excluded up to 2.6, 1.7 and 0.8 TeV through the processes $pp \to \tilde{g}\tilde{g} \to jets + E^{miss}_T$, $pp \to \tilde{t}\tilde{t}^* j \to j + E^{miss}_T$ and $ pp \to \tilde{\chi}^\pm_1\tilde{\chi}^0_2 \to \ell^+\ell^- + E^{miss}_T + jets$ for the gluino, stop and wino coannihilations at the $2\sigma$ level, respectively. While there is still no sensitivity of the neutralino DM through the process $ pp \to j\tilde{\tau}_1 \tilde{\tau}^*_1 \to j+\tau_h+E^{miss}_{T} $ in stau coannihilation at the HE-LHC, because of the low rate of the direct stau pair production and the soft tau from the decay of stau.

\section*{Acknowledgement}
Part of this work was done while M. A. was visiting Nanjing Normal University. This work was supported by the National Natural Science Foundation of China (NNSFC) under grant Nos. 11705093 and 11675242.


\end{document}